\documentclass[iop,apj,twocolappendix,numberedappendix]{emulateapj}
\usepackage[varg]{txfonts}
\usepackage{bm,color}

\newcommand{\beq}{\begin{equation}}
\newcommand{\eeq}{\end{equation}}
\newcommand{\beqn}{\begin{eqnarray}}
\newcommand{\eeqn}{\end{eqnarray}}
\newcommand{\pd}{\partial}
\newcommand{\AU}{{\rm AU}}

\newcommand{\eqref}[1]{(\ref{#1})}

\newcommand{\dint}{\displaystyle\int}
\newcommand{\dfrac}[2]{ {\displaystyle\frac{#1}{#2}} }
\newcommand{\tfrac}[2]{ {\textstyle\frac{#1}{#2}} }
\newcommand{\pfrac}[2]{ \biggl(\dfrac{#1}{#2}\biggr) }

\newcommand{\I}{{\rm I}}
\newcommand{\II}{{\rm II}}
\newcommand{\III}{{\rm III}}
\newcommand{\rin}{{r_{\rm in}}}
\newcommand{\rout}{{r_{\rm out}}}
\newcommand{\rrout}{r_{\rm out}^2}
\newcommand{\psiin}{{\psi_{\rm in}}}
\newcommand{\fin}{{f_{\rm in}}}
\newcommand{\fout}{{f_{\rm out}}}

\newcommand{\EllipticK}{{K}}
\newcommand{\EllipticE}{{E}}

\renewcommand{\leq}{\leqslant}
\renewcommand{\geq}{\geqslant}
\renewcommand{\apjl}{{ApJL}}

\renewcommand{\nat}{{Natur}}

\defcitealias{LPP94a}{LPP94a}
\defcitealias{TO14}{Paper~II}

\shorttitle{Radial Transport of Large-Scale Magnetic Fields in Accretion Disks. I.}
\shortauthors{Okuzumi, Takeuchi, \& Muto}
\slugcomment{ApJ, in press}

\begin{document}

\title{Radial Transport of Large-Scale Magnetic Fields in Accretion Disks. I. Steady Solutions and an Upper Limit on the Vertical Field Strength}

\author{Satoshi Okuzumi$^1$, Taku Takeuchi$^1$, and Takayuki Muto$^2$}
\affil{
$^1$Department of Earth and Planetary Sciences, Tokyo Institute of Technology, Meguro-ku, Tokyo 152-8551, Japan; okuzumi@geo.titech.ac.jp \\
$^2$Division of Liberal Arts, Kogakuin University, 1-24-2, Nishi-Shinjuku, Shinjuku-ku, Tokyo 163-8677, Japan}

\begin{abstract}
Large-scale magnetic fields are key ingredients of magnetically driven disk accretion. We study how large-scale poloidal fields evolve in accretion disks, with the primary aim of quantifying the viability of magnetic accretion mechanisms in protoplanetary disks. We employ a kinematic mean-field model for poloidal field transport and focus on steady states where inward advection of a field balances with outward diffusion due to effective resistivities. We analytically derive the steady-state radial distribution of poloidal fields in highly conducting accretion disks.  The analytic solution reveals an upper limit on the strength of large-scale vertical fields attainable in steady states. Any excess poloidal field will diffuse away within a finite time, and we demonstrate this with time-dependent numerical calculations of the mean-field equations. We apply this upper limit to large-scale vertical fields threading protoplanetary disks. We find that the maximum attainable strength is about $0.1~{\rm G}$ at $1~\AU$, and about $1~{\rm mG}$ at $10~\AU$ from the central star. When combined with recent magnetic accretion models, the maximum field strength translates into the maximum steady-state accretion rate of $\sim 10^{-7}~M_\odot~{\rm yr}^{-1}$, in agreement with observations. We also find that the maximum field strength is $\sim 1~{\rm kG}$ at the surface of the central star provided that the disk extends down to the stellar surface. This implies that any excess stellar poloidal field of strength $\ga {\rm kG}$ can be transported to the surrounding disk. This might in part resolve the magnetic flux problem in star formation.
\end{abstract}


\maketitle


\section{Introduction}
Young stars host gaseous accretion disks called protoplanetary disks.
They are a by-product of star formation from a molecular cloud.
Observed protoplanetary disks (of age $\sim 1~{\rm Myr}$) are 
characterized by an outer radius of $\sim 100~\AU$ \citep{K+02,A+09} 
and an accretion rate of $\sim 10^{-9}$--$10^{-7}~M_\odot~{\rm yr}^{-1}$ 
\citep{H+98,C+04,S-A+05}. 
To understand how the disks form, evolve, and dissipate with time 
is essential for understanding how planets form there. 
 
It is widely accepted that large-scale magnetic fields play key roles in disk evolution.
They not only drive turbulence via magnetorotational instability \citep[MRI;][]{BH91,BH98} but also accelerate winds and outflows via the magnetocentrifugal mechanism \citep{BP82,S96}. 
Both mechanisms transport the angular momentum of disks, leading to accretion of disk material.
{ In addition, turbulence driven by MRI has many effects on the motion of solid bodies in the disks 
and hence on planet formation. The effects include aerodynamical stirring of small dust particles {\citep[e.g.,][]{CSP05,TWBY06,FP06,JKM06,OH11,OH12}} and gravitational stirring of larger solid bodies \citep[e.g.,][]{LSA04,NP04,N05,YMM09,YMM12,GNT11,GNT12,OO13a,OO13b}.}

A key parameter of these accretion mechanisms is 
the strength of the poloidal field threading the disk.
Early local-box MHD simulations  by \citet{HGB95} 
already suggested a positive dependence of 
the MRI-driven accretion stress on the large-scale field strength
(see also \citealt*{SITS04,PCP07}).
More realistic simulations including 
vertical gas stratification \citep{SMI10,BS13a},
ohmic diffusion \citep{OH11,GNT12},
and/or ambipolar diffusion \citep{S+13a,S+13b} have not changed this basic picture in the sense that the vertical average of the accretion stress increases 
with the strength of the net (large-scale) vertical field.  
The accretion stress produced by magnetocentrifugal winds also exhibits 
a positive dependence on the net poloidal field strength \citep{S+13b,BS13a,BS13b,B13}.  
Accretion disk models taking into account these dependences predict 
that the net vertical field strength determines the fate of disk evolution \citep{SMI10,ASM13}.
Therefore, in order to understand the evolution of accretion disks, 
one needs to understand the evolution of poloidal fields threading the disks.

Radial transport of a large-scale poloidal field is a long-standing issue in the theory of accretion disks.
Early studies argued that accretion disks would advect weak interstellar magnetic field inward and build up a strong field at the center of the disks \citep{BR74,L76}. However, this picture was later confronted with an issue raised by \citet{vB89} and \citet[][henceforth \citetalias{LPP94a}]{LPP94a}. They argued that if turbulence is the source of disk accretion, 
then large-scale fields would diffuse away faster than they are advected inward. 
Their argument is as follows. Consider a turbulent disk and assume that the turbulence acts as an effective viscosity $\nu_{\rm turb}$ on disk matter and as an effective resistivity (diffusivity) $\eta_{\rm turb}$ on large-scale magnetic fields. Then, an { order-of-magnitude estimate} of the mean-field induction equation shows that significant field dragging occurs only when ${\cal D} \equiv ({\rm Pm}_{\rm turb} h)^{-1}$ is less than unity, where $h$ is the disk's aspect ratio ($\ll 1$ for thin disks) and ${\rm Pm}_{\rm turb} = \nu_{\rm turb}/\eta_{\rm turb}$ is the turbulent magnetic Prandtl number. The large factor $h^{-1}$ in ${\cal D}$ comes from  
the fact that  even a slightly bent poloidal field can lead to a large electric current in a thin accretion disk.
One finds that if ${\rm Pm}_{\rm turb} \sim 1$, as is the case for MRI turbulence \citep{GG09,LL09,FS09}, 
then the above criterion is violated, i.e., diffusion would prevent advection of the field. 
This has been invoked as a challenge to the viability of magnetically driven jets and winds.
This also implies that MRI turbulence would saturate at a low level. 
For protoplanetary disks, the level might be too low to be consistent with observations \citep{BS13b,S+13a}.

However, recent studies suggest that poloidal fields are more likely 
to be advected than previously thought. 
\cite{SU05} point out that turbulent diffusion is reduced if the magnetic flux 
passes through the disk in concentrated patches.
Turbulent diffusion can also be prevented if the disk 
has highly conductive (non-turbulent) surface layers {\citep{BL07,BL12,RL08,LRB09}}. 
\citet{GO12,GO13} point out that the advection of magnetic flux can be significantly faster 
than that of mass owing to fast radial velocities in the low-density regions away 
from the midplane \citep{TL02}. 
{ Efficient advection of magnetic flux is also suggested by 
a number of global MHD simulations \citep{INA03,BHK09,TNM11,SI14},
some of which \citep{BHK09,SI14} indicate that the advection indeed takes place
at large distances from the midplane.}
Taken together, these results suggest that poloidal fields can be efficiently dragged inward 
(i.e., ${\cal D}$ can be $ < 1$) in a realistic accretion disk.

These recent studies motivate us to ask the following questions. 
Assuming that advection of a poloidal field is indeed efficient, 
how will the radial distribution of the poloidal field evolve with time? 
Can we constrain the range of the poloidal field strength for such a conducting  accretion disk?  What are the implications for magnetically driven evolution of accretion disks such as protoplanetary disks? 
Aiming at addressing these questions, 
we revisit global transport of large-scale magnetic fields
with the assumption that poloidal fields can be efficiently dragged inward 
in accretion disks. 
As a first step of this project, we here focus on steady states 
where inward advection of a magnetic flux balances with outward diffusion.
We employ the kinematic mean-field formalism of \citetalias{LPP94a} and 
analytically derive the steady-state distribution of a poloidal field 
for highly conducting accretion disks.
We find that  there exists an upper limit on the vertical field strength
for a given distance from the central star.
When combined with recent models for MRI and wind-driven accretion, 
this upper limit suggests that magnetically driven accretion does not 
produce an accretion rate higher than $\sim 10^{-7}~M_\odot~{\rm yr^{-1}}$
as long as the magnetic field configuration reaches a steady state.
A companion paper addresses non-steady field transport 
in evolving viscous accretion disks  \citep{TO14}.

The plan of the paper is as follows. In Section~\ref{sec:basic} we introduce the model to describe radial transport of magnetic fields, and derive the condition 
to be satisfied in steady states. In Section~\ref{sec:analytic} we analytically derive
the steady-state equations as well as the upper limit on the vertical field strength. Section~\ref{sec:comp} compares the analytic solution with time-dependent numerical solutions. In Section~\ref{sec:application} we apply our result to protoplanetary 
systems and discuss its astrophysical implications.
Section~\ref{sec:summary} is devoted to a summary.    
    
\section{Model}\label{sec:basic}
\begin{figure}[t]
\epsscale{1.15}
\plotone{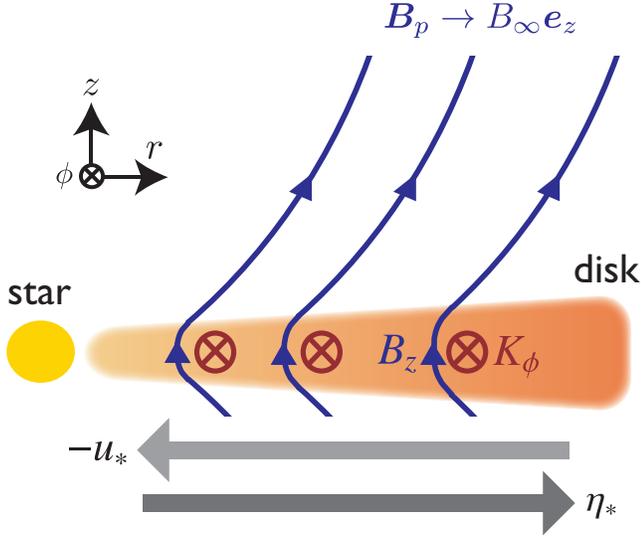}
\caption{Schematic illustration of the mean-field model adopted in this study. 
The disk is assumed to be threaded by a large-scale poloidal field 
${\bm B}_p = B_r {\bm e}_r + B_z {\bm e}_z$ (solid curves; note that the field lines below the disk are omitted). The field connects to a uniform field $B_\infty{\bm e}_z$ at infinity. 
The flux is advected inward at some averaged speed $|u_*|$ (Equation~\eqref{eq:ustar}) and is diffused outward by some effective resistivity $\eta_*$ (Equation~\eqref{eq:etastar}).
The surface density of the toroidal current, $K_\phi$, is related to the flux distribution on the disk by Biot--Savart's law (Equation~\eqref{eq:BS}).
}
\label{fig:setting}
\end{figure}

In this study, we adopt a mean-field model for large-scale poloidal fields 
developed by \citetalias{LPP94a} (see Figure~\ref{fig:setting} for a schematic illustration).
We take a cylindrical coordinate system ($r$, $\phi$, $z$) 
with $r = 0$ at the disk center and $z=0$ on the disk midplane.
We denote the large-scale averages of the magnetic field strength 
and neutral gas velocity as ${\bm B}$ and ${\bm u}$, respectively.
The adopted model is kinematic in the sense that ${\bm u}$ is given 
as a prescribed function. Any small-scale (or turbulent) motion 
in the disk is treated as a macroscopic magnetic diffusivity, 
which will also be given as a prescribed quantity (see Section~\ref{sec:diskfield} below).
In principle, one can evolve ${\bm B}$ and ${\bm u}$ in a self-consistent way
by solving dynamical equations or adopting any empirical relation between 
${\bm B}$ and ${\bm u}$ \citep[see][]{AP96},
but we defer this to future work.

We assume that both ${\bm B}$ and ${\bm u}$ are axisymmetric, or that 
these quantities are already averaged in the azimuthal direction.
Therefore, we do not explicitly treat transport of strongly magnetized patches as considered by \citet{SU05}.
We neglect conversion of toroidal into poloidal fields by turbulence (the so-called $\alpha$ dynamo),
though such effects do exist in MRI-driven turbulence \citep[e.g.,][]{B+95,DSP10,G10,SHB11,F+12a,F+12b}.
With this assumption together with axisymmetry, the toroidal field $B_\phi$ does not
enter the induction equation for the poloidal field. 
For this reason, we will treat the disk field as purely poloidal.

The accretion disk is assumed to be geometrically thin with half thickness $H \ll r$.
The large-scale electric current ${\bm J} = (c/4\pi)\nabla\times {\bm B}$ 
is purely toroidal since ${\bm B}$ is axisymmetric and poloidal. We assume that the toroidal current is well confined to the disk ($|z|<H$). The magnetic field exterior to the disk is therefore a potential field. 

We express the poloidal field in terms of a flux function $\psi$ defined as 
${\bm B} = \nabla\times (\psi {\bm e}_\phi/r)$. 
By axisymmetry, the radial and azimuthal components of the field are written as 
\beq
B_r = -\frac{1}{r}\frac{\pd \psi}{\pd z},
\label{eq:Br}
\eeq
\beq
B_z = \frac{1}{r}\frac{\pd \psi}{\pd r}.
\label{eq:Bz}
\eeq
The flux function is proportional to the magnetic flux $\Phi(r)$
passing through the disk interior to $r$, i.e., 
\beq
\Phi(r) = 2\pi \int_0^r B_z(r',0) r' dr' = 2\pi \psi(r,0).
\eeq

\subsection{Disk-field Equation}\label{sec:diskfield}

The time evolution of a poloidal flux distribution 
is determined by the mean-field induction equation. 
In terms of $\psi$, this can be written as (see Equation~(10) of \citetalias{LPP94a}) 
\beq
\frac{\pd \psi}{\pd t} =  - u_r \frac{\pd \psi}{\pd r} - \frac{4\pi r\eta}{c} J_\phi,
\label{eq:psi}
\eeq
where $u_r(r,z)$ is the mean radial velocity, $\eta(r,z)$ is the magnetic diffusivity (resistivity), and 
\beq
J_\phi = \frac{c}{4\pi}(\nabla\times{\bm B})_\phi 
= \frac{c}{4\pi}\left(\frac{\pd B_r}{\pd z} - \frac{\pd B_z}{\pd r}\right)
\label{eq:Jphi}
\eeq 
is the toroidal component of the mean electric current.
In the framework of the mean-field theory, the diffusivity in Equation~\eqref{eq:psi} may be interpreted as the sum of the molecular (microscopic) diffusivity $\eta_{\rm mol}$ and the turbulent (macroscopic) diffusivity $\eta_{\rm turb}$.
Formally, Equation~\eqref{eq:psi} neglects the Hall term and ambipolar diffusion. 
However,  if  ${\bm B}$ is poloidal and ${\bm J}$ is toroidal as envisaged here, 
then ambipolar diffusion is identical to ohmic diffusion (because ${\bm J}\perp{\bm B}$), and the toroidal component of the Hall term vanishes (because the Hall term $\propto {\bm J}\times{\bm B}$).\footnote{We note that this argument holds only when the cross-correlations of small-scale quantities arising from the non-ohmic terms are negligible. For example, see Equation~(24) of \citet{KL13} for a mean-field equation including a cross-correlation from the Hall term. { We also note that accretion disks can in fact have a strong toroidal field owing  to the orbital shear \citep{TS08}.} 
} In this case, the non-ohmic effects only amount to adding the ambipolar diffusivity to $\eta$ 
in Equation~\eqref{eq:psi} \citep[the so-called Pedersen diffusivity; see, e.g.,][]{WS12}.

For a geometrically thin disk, where $\psi$ is approximately independent of $z$, 
one can reduce Equation~\eqref{eq:psi} to a one-dimensional equation 
with any vertical averaging (\citetalias{LPP94a}, \citealt*{OL01}).
In this study, we follow \citet{OL01} and average 
the equation after weighting the ``conductivity'' $\eta^{-1}$. 
Dividing Equation~\eqref{eq:psi} by $\eta$ and integrating it over $-H<z<H$, we obtain the conductivity-weighted average of the induction equation (see Equation (9) of \citealt*{OL01})
\beq
\frac{\pd \psi}{\pd t} = -u_* \frac{\pd \psi}{\pd r} 
- \frac{2\pi r\eta_*}{c H} K_\phi,
\label{eq:psi_avr}
\eeq
where
\beq
K_\phi(r) \equiv  \int_{-H}^H J_\phi(r,z) dz
\label{eq:Kphi}
\eeq
is the toroidal component of the surface current density, 
\beq
u_*(r) 
\equiv \frac{\eta_*}{2H} \int_{-H}^{H} \frac{u_r(r,z)}{\eta(r,z)}dz
\label{eq:ustar}
\eeq
is the conductivity-weighted average of the radial gas velocity, and
\beq
\frac{1}{\eta_*(r)} 
\equiv \frac{1}{2H} \dint_{-H}^H\dfrac{dz}{\eta(r,z)}
\label{eq:etastar}
\eeq
is the height average of the conductivity.
An advantage of the conductivity-weighted average is that 
the resulting induction equation (Equation~\eqref{eq:psi_avr}) 
does not involve the vertical distribution of $J_\phi$.
The vertically integrated current $K_\phi$ is sufficient, 
and this can be determined from the field configuration exterior to the disk 
 (\citealt*{OL01}; see also Section~\ref{sec:BS} below). 
Furthermore, it naturally accounts for the fact that
inward advection of the flux mainly occurs at heights where the conductivity 
$\eta^{-1}$ is high. 
This becomes important when an upper layer of the disk has 
a high conductivity \citep{BL07,RL08} or a high accretion velocity \citep{GO12,GO13}.

\subsection{Exterior-field Equation}\label{sec:BS}
Equation~\eqref{eq:psi_avr} requires a relation between $\psi$ and 
the surface current density $K_\phi$. 
In this study, we follow \citetalias{LPP94a} and 
determine $K_\phi$ from Biot--Savart's law.
We assume that the current vanishes outside the disk 
(i.e., the exterior magnetic field is a potential field), 
and that the exterior field asymptotically approaches 
at infinity a uniform vertical field of strength $B_\infty$. 
Under these assumptions, the flux function $\psi$ can be decomposed as
(see Equation (17) of \citetalias{LPP94a})
\beq
\psi = \psi_\infty + \psi_d,
\label{eq:psi_decomp}
\eeq
where
\beq
\psi_\infty(r) = \frac{1}{2}B_\infty r^2
\label{eq:psi_infty}
\eeq
is the contribution from the imposed field and
$\psi_d$ is from the induced field.
The latter is related to $K_\phi$ via Biot--Savart's law \citepalias{LPP94a}\footnote{As shown in Appendix~\ref{sec:BS_append}, Equation~\eqref{eq:BS} is equivalent to 
Equation (18) of \citetalias{LPP94a} \citep[see also][]{O97}.}
\beq
\psi_d(r) = \frac{4}{c}\int
 R(r_<,r_>) K_\phi(r')dr',
\label{eq:BS}
\eeq
\beq
R(r_<,r_>) \equiv r_>\left[\EllipticK\pfrac{r_<}{r_>}-\EllipticE\pfrac{r_<}{r_>}\right],
\eeq
where $r_< = \min\{r,r'\}$, $r_> = \max\{r,r'\}$, and
$\EllipticK (x)$ and $\EllipticE (x)$ are 
the complete elliptic integrals defined by
$\EllipticK(x) =\int_0^{\pi/2}(1-x^2\sin^2\theta)^{-1/2}d\theta$
and $\EllipticE(x) = \int_0^{\pi/2}(1-x^2\sin^2\theta)^{1/2}d\theta$.
The set of Equations~\eqref{eq:psi_avr} and \eqref{eq:BS} determines 
the radial transport of the net poloidal flux 
in a disk under the assumptions we have employed. 

\subsection{Steady-state Condition}\label{sec:steady}
In this study, we focus on the steady-state solutions 
of Equations~\eqref{eq:psi_avr} and~\eqref{eq:BS}.
For a steady state ($\pd\psi/\pd t = 0$), 
Equation~\eqref{eq:psi_avr} gives the relation 
between $B_z$ and $K_\phi$,
\beq
B_z =  \frac{2\pi D}{c}K_\phi,
\label{eq:staticcond}
\eeq
where the dimensionless coefficient $D$ is defined as
\beq
D \equiv 
-\frac{\eta_*}{u_*H} = -
\frac{2}{\int_{-H}^H (u_r/\eta) dz}.
\label{eq:D}
\eeq
Note that we have rewritten $d\psi/dr$ in terms of $B_z$
using Equation~\eqref{eq:Bz}. 

Equation~\eqref{eq:staticcond} determines the bending angle 
of the poloidal field on the disk surface as a function of $D$.
Substituting Equation~\eqref{eq:Jphi} into Equation~\eqref{eq:Kphi}
and assuming that the poloidal field has dipolar symmetry,
we have
\beq
K_\phi = \frac{c}{2\pi} B_{rs} 
-\frac{c}{4\pi} \frac{\pd}{\pd r}\int_{-H}^{H} B_z dz,
\label{eq:Kphi_Brs}
\eeq
where $B_{rs} \equiv B_r|_{z=H} (= - B_r|_{z=-H})$
is the radial field strength on the disk surface.
If $|B_{rs}/B_z| \gg H/r$, the second term on the right-hand side 
of Equation~\eqref{eq:Kphi_Brs} is negligible, so 
Equation~\eqref{eq:staticcond} approximately gives
\beq
\frac{B_{rs}}{B_z} \approx \frac{1}{D}.
\label{eq:Bratio_static}
\eeq
This equation implies that the poloidal field lines are bent 
by angle $i$ from the vertical such that $\tan i \approx 1/D$.
This well explains the numerical findings of \citetalias{LPP94a} 
showing that $\tan i = 1.52/(3D/2) = 1.01/D$ (see their Equation (40); note that ${\cal D}$ of \citetalias{LPP94a} differs from our $D$ by factor $2/3$).

Equation~\eqref{eq:staticcond} does not solely determine 
how each of $B_{rs}$ and $B_z$ depends on $r$.
In order to know this, one needs to solve Biot--Savart's equation
(Equation~\eqref{eq:BS}) simultaneously with Equation~\eqref{eq:staticcond}.
Section~\ref{sec:analytic} will be devoted to this task.

\subsection{Assumption about $D$ and Definition of Regions}
\begin{figure}[t]
\epsscale{1.15}
\plotone{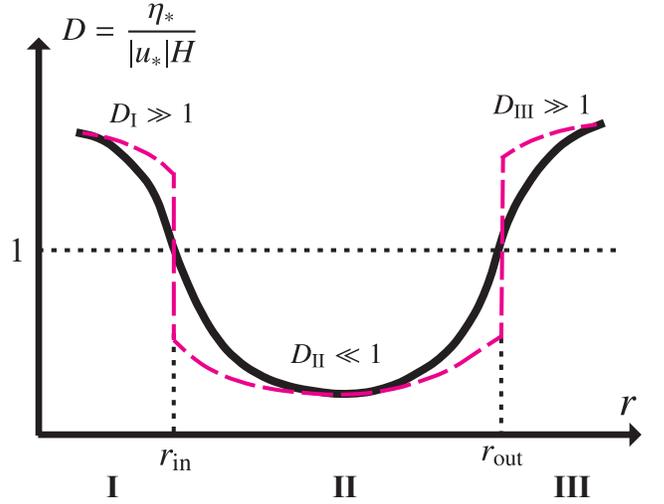}
\caption{Schematic radial profile of $D$ assumed in this study (solid curve). 
The disk is highly advecting ($D \ll 1$) in the bulk of the disk (region II),
with diffusion being dominant ($D \gg 1$) only in the innermost and outermost parts of the disk (regions I and III).  The three regions are defined by the boundaries where $D$ crosses unity, $r= \rin$ and $\rout$. The analytic steady solution derived in Section~\ref{sec:analytic} approximates the boundaries as sharp transitions as depicted by the dashed curve. Smooth transitions like the solid curve will be considered in numerical calculations (Section~\ref{sec:smoothD}).
}
\label{fig:D}
\end{figure}

As mentioned earlier, recent theoretical studies have suggested 
that inward dragging of a poloidal field can efficiently take place 
in realistic accretion disks. 
We seek to understand how then a poloidal magnetic flux 
would be distributed in such a conductive disk.
Unfortunately, the value of $D$ is highly dependent on 
the vertical structure of an accretion disk, which is yet to be understood quantitatively. 
Therefore, in this study, we employ a simply toy model for 
the radial distribution of $D$ as illustrated in Figure~\ref{fig:D}. 
Here, it is assumed that inward advection dominates over outward diffusion (i.e., $D<1$) at $\rin < r < \rout$, and that the opposite happens (i.e., $D>1$) at $r<\rin$ and $r>\rout$. 
We will refer to the three regions as region I, II, and III from inside to outside (see Figure~\ref{fig:D}), 
and the quantities in different regions will be distinguished with subscripts ``I,'' ``II,'' and ``III.'' 
Region II corresponds to the body of a highly conducting disk,  
while region III may be regarded as the disk's outer edge where the 
ambipolar diffusivity is high ($\eta_* \to \infty$) as suggested by \citet{D+13}.
{ Region I may be considered as the central star where the accretion of matter terminates ($u_* \to 0$). 
In reality, there can be a magnetosphere and a jet near the interface of the star and disk \citep{S+94},
but inclusion of these complexities is deferred to future work.}

\section{Analytic Steady Solution}\label{sec:analytic} 
In this section, we derive the steady-state distribution 
of the poloidal field for the assumed  distribution of $D$.
We do this by solving Equations~\eqref{eq:staticcond} and \eqref{eq:BS} analytically.

To make this problem analytically tractable, 
we for the moment approximate $D$ as a piecewise smooth function 
with sharp transitions at $r = \rin$ and $r = \rout$ is sharp (see the dashed line in Figure~\ref{fig:D}).
Thus, in the following analysis, $D$ is either $\gg 1$ or $\ll 1$.
The resulting analytic solution, however, approximates the solution for smooth $D$ 
in a surprisingly good accuracy, as we will see in Section~\ref{sec:smoothD}.
In addition, we assume that the surface currents 
in regions I and III are so small that $\psi_d$ is 
predominantly determined by the surface current in region II.
Thus, we limit the interval of integration in Equation~\eqref{eq:BS} 
to region II,  $\rin < r' <  \rout $.
This is a good assumption since the surface current is generally 
suppressed in highly diffusive ($D \gg 1$) regions.


The outline of our analysis is as follows. 
In Section~\ref{sec:perturb}, we expand variables and equations 
in powers of $D_\II (\ll 1)$. This allows us to solve Equation~\eqref{eq:staticcond} by successive substitution. We then identify in Section~\ref{sec:asympt} how the leading-order quantities  behave away from the boundaries ($\rin \ll r \ll \rout$). We use the results to infer and derive the full solution for all regions, which will be done in Sections~\ref{sec:Kphi} and \ref{sec:Bz}.
Readers who are primarily interested in the final form of the solution and its implication can skip to Sections~\ref{sec:Bz} and~\ref{sec:Bzmax}.

\subsection{Perturbative Expansion in Powers of $D_\II$}\label{sec:perturb}
Given $D_{\rm II} \ll 1$, we are allowed to expand all the quantities and equations in powers of $D_{\rm II}$ and solve the equations perturbatively.
Let us expand $B_{z,\II}$ and $K_{\phi,\II}$ as
\beq
B_{z,\II} = B_{z,\II}^{(0)} + B_{z,\II}^{(1)} + O(D_\II^2), 
\label{eq:Bz_eps}
\eeq
\beq
K_{\phi,\II} = K_{\phi,\II}^{(0)} + O(D_\II), 
\label{eq:Kphi_eps}
\eeq
where $B_{z,\II}^{(n)}, K_{\phi,\II}^{(n)} = O(D_\II^n)$ ($n=0,1,\dots$), and similarly for all other variables in all regions.

Substituting Equations~\eqref{eq:Bz_eps} and \eqref{eq:Kphi_eps} 
 into Equation~\eqref{eq:staticcond} 
and collecting terms involving the same power of $D_\II$, 
we obtain
\beq
B_{z,\II}^{(0)} = 0,
\label{eq:staticcond_0}
\eeq
\beq
B_{z,\II}^{(1)} = \frac{2\pi D_\II(r)}{c} K_{\phi,\II}^{(0)}
\label{eq:staticcond_1}
\eeq
to first order in $D_\II$. 
Equation~\eqref{eq:staticcond_0} means that 
region II is completely devoid of vertical fields in the limit of $D_\II \to 0$. 
The vertical field strength is nonzero only to first order in $D_\II$, and it is determined by the {\it zeroth-order} surface current $K_{\phi,\II}^{(0)}$.
It follows from Equations~\eqref{eq:Bz} and \eqref{eq:staticcond_0} 
that $\psi_\II^{(0)}$ is independent of $r$ and its value is determined by the flux inside region I, namely,
\beq
\psi_\II^{(0)} =  \psi(\rin) \equiv \psiin,
\eeq
where we have denoted the constant as $\psiin$.
From this and Equation~\eqref{eq:psi_decomp}, the disk-induced component 
of $\psi_{\II}^{(0)}$ is a quadratic function of $r$,
\beq
\psi_{d,\II}^{(0)}(r) = \psiin - \psi_\infty(r)
=  \psiin -\frac{1}{2}B_\infty r^2.
\label{eq:psid_II}
\eeq
This means that to zeroth order in $D_\II$, 
the disk-induced field ($r^{-1} \pd \psi_{d,\II}/\pd r$) in region II 
exactly cancels the imposed field ($B_z$).
Substituting \eqref{eq:psid_II} into the Biot--Savart equation (Equation~\eqref{eq:BS}), 
we obtain the equation for $K_{\phi,\II}^{(0)}$,
\beq
\psiin - \psi_\infty(r)
= \frac{4}{c} \int_\rin^\rout R(r_<,r_>)K_{\phi,\II}^{(0)}(r')dr',
\label{eq:BS_0}
\eeq
where $\rin < r < \rout$.
Note that we have neglected the current in regions I and III, 
as we already stated at the beginning of this section. 
The following subsections will be devoted to solving this equation.

\subsection{Asymptotic Solution Deep Inside Region II}\label{sec:asympt}
It is still not an easy task to find the solution to Equation~\eqref{eq:BS_0}.
Therefore, it is useful to see how $K_{\phi,\II}^{(0)}$ behaves far away from the inner and outer boundaries. To do this, let us for the moment take the limits of $\rin \to 0$ and $\rout \to \infty$. We also temporarily drop $\psi_\infty(r)$ on the left-hand side of Equation~\eqref{eq:BS_0} assuming that the induced component $\psi_d$ dominates the total flux deep inside region II (this assumption will be validated a posteriori in Section~\ref{sec:Kphi}). Under these simplifications, Equation~\eqref{eq:BS_0} reduces to 
\beq
\psiin \approx \frac{4}{c} \int_0^\infty R(r_<,r_>)K_{\phi,\II}^{(0)}(r')dr'.
\label{eq:BS_0_hom}
\eeq

The solution to Equation~\eqref{eq:BS_0_hom} can be easily found 
by assuming $K_{\phi,\II}^{(0)}$ of the power law from
\beq
K_\phi^{(0)}(r) = A r^{-2},
\label{eq:Kphi_hom}
\eeq
where $A$ is a constant. For this $K_{\phi,\II}^{(0)}$, 
the right-hand side of Equation~\eqref{eq:BS_0_hom} indeed becomes 
a constant, 
\beqn
&&\frac{4A}{c} \int_0^\infty
R(r_<,r_>) r'^{-2} dr' 
\nonumber \\
&& = \frac{4A}{c}  \int_0^{1} \left[\EllipticK(x) - \EllipticE(x)\right](x^{-2} + x^{-1})dx
\nonumber \\
&& = \frac{2\pi A}{c},
\eeqn
where we have used that 
$ \int_0^{1} \left[\EllipticK(x) - \EllipticE(x)\right](x^{-2} + x^{-1})dx
= \pi/2$ (this can be proven with {\it Mathematica}).
Comparing this with the left-hand side of Equation~\eqref{eq:BS_0_hom}, 
we find 
\beq
A = \frac{c\psiin}{2\pi}.
\eeq
Consequently, we find that $K_{\phi,\II}^{(0)}$ at $\rin \ll r \ll \rout$ 
asymptotically behaves as 
\beq
K_{\phi,\II}^{(0)}(r) \approx \frac{c\psiin}{2\pi r^2}.
\label{eq:Kphi_hom_sol}
\eeq

\begin{figure}[t]
\epsscale{1.15}
\plotone{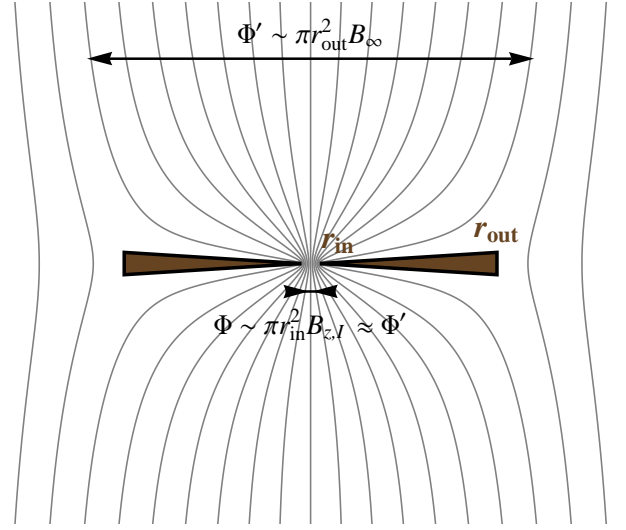}
\caption{Split-monopole configuration of the magnetic field above and below a highly conductive ($D \ll 1$) accretion disk. The gray wedges schematically show  an edge-on view of the disk, and the thin solid curves are the exterior field lines. Away from the disk, the exterior field approaches the unidirectional imposed field $B_\infty$. At $\sqrt{r^2+z^2} \ll \rout$, it approaches a monopolar field with the opposite polarity with respect to the equatorial plane.
{ The magnetic flux threading at $r \la \rin$ is approximately equal to the flux of the {\it imposed} field threading at $r \la \rout$ (i.e., $\Phi \approx \Phi'$; see Equation~\eqref{eq:Bz_I_asympt}).}
}
\label{fig:split}
\end{figure}
Equation~\eqref{eq:Kphi_hom_sol} implies that the exterior field has a split-monopole geometry (see Figure~\ref{fig:split}).
We recall that $K_\phi \approx (c/2\pi)B_{rs}$ when $B_{rs}/B_z \gg H/r$ (see Section~\ref{sec:steady}). Thus, Equation~\eqref{eq:Kphi_hom_sol} means that $B_{rs} = \psiin/r^2$ when  $D_\II \ll 1$. With this boundary condition and dipolar symmetry, potential theory tells us that the external field is a monopole field
of strength $|{\bm B}| = \psiin/(r^2+z^2)$ except that the field above the disk has the opposite sign to the field below. 
The split-monopole configuration is a natural consequence of flux accumulation at the center by a highly conductive ($D \ll 1$) accretion disk \citep[e.g.,][]{GS93,LS97,ASL03,G+06}.

From Equation~\eqref{eq:Kphi_hom_sol} and the steady-state condition (Equation~\eqref{eq:staticcond_1}), the asymptotic form of $B_{z,\II}^{(1)}$ is
\beq
B_{z,\II}^{(1)}(r) 
\approx \frac{D_\II(r)\psiin}{r^2}.
\label{eq:Bz_1}
\eeq
Note that $B_{z,\II}^{(1)}$ does not necessarily obey a power law 
since $D_\II$ is an arbitrary function of $r$.

It is also possible to find the asymptotic solution in region II for arbitrary values of  $D_\II$ as long as it is a constant. We defer the demonstration of this to Appendix~\ref{sec:constantD} because we will not use the results in the following analysis.

\subsection{Determination of $\psiin$}\label{sec:Kphi}

The asymptotic solution derived in Section~\ref{sec:asympt} 
involves $\psiin$, which is as yet undetermined. 
As we will show in this subsection, 
one can determine $\psiin$ by solving Equation~\eqref{eq:BS_0} 
taking into account the presence of the imposed field, $\psi_\infty$.

Let us relax the condition $r \ll \rout$ and instead 
assume  $K_{\phi,\II}^{(0)}$ of the form
\beq
K_{\phi,\II}^{(0)}(r) =  \frac{c\psiin}{2\pi r^2}\fout(r).
\label{eq:Kphi_inhom}
\eeq
Here the correction factor $\fout(r)$ satisfies $\fout \to 1$ at $r \ll \rout$ so that Equation~\eqref{eq:Kphi_inhom} reduces to Equation~\eqref{eq:Kphi_hom} there. The equation to be solved is 
\beq
\psiin - \frac{1}{2}B_\infty r^2 = \frac{2\psiin }{\pi}\int_0^\rout
R(r_<,r_>) \frac{\fout(r')}{r'^2} dr'.
\label{eq:BS_II}
\eeq

As we will show in Appendix~\ref{sec:out}, 
the requirement that the right-hand side of Equation~\eqref{eq:BS_II} 
be a quadratic function of $r$ specifies $\fout$ and $\psiin$.
We defer the derivation of it to the appendix and here only show the result.
We find that $\fout$ is well approximated by 
\beq
\fout(r) = 1 + \frac{r}{\rout} \left\{
\left[1-\pfrac{r}{\rout}^2 \right]^\gamma - 1
\right\},
\label{eq:fout}
\eeq
where $\gamma = 0.45$ is the best-fit parameter. 
For this $\fout$, the right-hand side of 
Equation~\eqref{eq:BS_II} becomes a quadratic form 
\beq
\frac{2\psiin}{\pi} \int_0^{\rout} R(r_<,r_>) K_{\phi,\II}^{(0)}(r')dr'
= 
\psiin \left[1-\frac{1}{2} \pfrac{r}{\rout}^2 \right].
\eeq
Finally, comparing this with the left-hand side of Equation~\eqref{eq:BS_II}, we find 
\beq
\psiin = B_\infty \rrout.
\label{eq:psiin}
\eeq
It follows from Equation~\eqref{eq:psiin} that $\psi_\infty(r) \ll \psiin$ at $r \ll \rout$, which validates the assumption we made in Section~\ref{sec:asympt}.

Comparison between Equations~\eqref{eq:psi_infty} and \eqref{eq:psiin} 
reveals an interesting relation 
\beq
\psiin = 2\psi_\infty(\rout).
\label{eq:psiin2}
\eeq 
In other words, the magnetic flux threading region I is 
exactly twice the flux of the {\it imposed} field threading regions I and II.
One half of $\psiin$ simply means that the highly advecting region II 
flushes out all the imposed flux toward region I. 
Another half of $\psiin$ comes from region III: advection and diffusion near $r = \rout$ do transport some fraction of the imposed flux in region III to region II.
An important point here is that the latter process is however limited: 
even a highly conducting disk cannot convey an arbitrarily large poloidal flux 
from region III to region I.
As we will see in Section~\ref{sec:Bz}, this fact sets an upper limit on the vertical field strength in region I.

For later convenience, we will also show how Equation~\eqref{eq:Kphi_hom} should be corrected near the inner boundary $r=\rin$. This correction becomes important when evaluating the magnetic flux inside region I (see Section~\ref{sec:Bz}). 
Let us assume $K_{\phi,\II}^{(0)} = (c\psiin/2\pi r^2) \fin(r)$, where $\fin$ satisfies $\fin \to 1$ at $r \gg r_{\rm in}$. Substituting this into Equation~\eqref{eq:BS_0} and assuming $\rin \sim r \ll \rout$, we obtain the equation for $\fin$, 
\beq
\psiin = \frac{2\psiin}{\pi}\int_\rin^\infty R(r_<, r_>) \frac{\fin(r')}{r'^{2}} dr'.
\eeq
The left-hand side of this equation requires that 
the right-hand side be independent of $r$.
As shown in Appendix~\ref{sec:out}, this requirement specifies $\fin$  as
\beq
\fin(r) = \left[ 1 - \Bigl(\frac{r}{\rin}\Bigr)^{-2} \right]^{-1/2} dr'.
\label{eq:fin}
\eeq
Note that $\fin$ diverges at $r = \rin$, but the radially integrated current 
$I_\phi = 2\pi \int r K_\phi(r) dr$ is still finite.
The singularity merely reflects the sharp transition of the assumed $D$ at $r = \rin$.

\subsection{The Final Form of the Solution and an Example}\label{sec:Bz}

To summarize the previous subsection, we have found that
the surface current density in region II is given by 
\beq
K_{\phi,\II}(r) = \frac{c B_\infty}{2\pi} \pfrac{\rout}{r}^2
\fin(r) \fout(r). 
\label{eq:Kphi_II_sol}
\eeq
This expression is correct to zeroth order in $D_\II$ (we will omit the superscripts ``(0)'' and ``(1)'' from this subsection onward).
The correction factors $\fin$ and $\fout$ are given by 
Equations~\eqref{eq:fin} and \eqref{eq:fout}, respectively. These factors are only important near the inner and outer boundaries of region II, i.e., $\fin \approx 1$ at $r \gg \rin$ and $\fout \approx 1$ at $r \ll \rin$.

We are now able to derive the radial distribution of the induced field for all regions.
For region II, Equations~\eqref{eq:Kphi_II_sol} and \eqref{eq:staticcond_1} give
\beq
B_{z,\II} = D_\II \pfrac{\rout}{r}^2 \fin(r)\fout(r) B_\infty ,
\label{eq:Bz_II_sol}
\eeq
to first order in $D_\II$. 
For regions I and III, the Biot--Savart equation (Equation~\eqref{eq:BS})
directly determines the induced flux as 
\beq
\psi_{d,\I}(r) = \frac{4}{c}\int_\rin^\rout R(r,r')K_{\phi,\II} dr',
\label{eq:BS_0_I}
\eeq
\beq
\psi_{d,\III}(r) = \frac{4}{c}\int_\rin^\rout R(r',r)K_{\phi,\II} dr',
\label{eq:BS_0_III}
\eeq
where we have used that $(r_<,r_>) = (r,r')$ for region I and 
$(r_<,r_>) = (r',r)$ for region III.
In Appendix~\ref{sec:C}, we perform these integrations 
in an exact way and derive the analytic expressions for $\psi_{d,\I}$ and $\psi_{d,\III}$.
These are given by Equations~\eqref{eq:psid_I_exact} and \eqref{eq:psid_III_exact}, 
with the corresponding field strengths by Equations~\eqref{eq:Bz_I_exact} and \eqref{eq:Bzd_III_exact}, respectively.

We are particularly interested in the vertical field strength in region I.
Although its exact expression is given in Appendix~\ref{sec:C} (Equation~\eqref{eq:Bz_I_exact}), it is more instructive to derive its asymptotic expression at $r \ll \rin$ directly from Equation~\eqref{eq:BS_0_I}. 
For $r \ll \rin$, we may approximate the kernel $R$ as $R(r,r') \approx \pi r^2/4r'$, which follows from the asymptotic expansion of the elliptic integrals.
We may also take $\rout \to \infty$ and hence $\fout \approx 1$ 
because the current at $r \sim \rout$ has little effect on 
the magnetic flux at $r \ll \rin$. 
Using these approximations and Equations~\eqref{eq:Kphi_II_sol} and~\eqref{eq:BS_0_I}, we obtain 
\beqn
\psi_{d,\I}(r) &\approx&  \frac{1}{2}B_\infty \rrout r^2 
\int_{\rin}^\infty \frac{\fin(r')}{r'^3} dr'
\nonumber \\
&=& \frac{1}{2}\pfrac{\rout}{\rin}^2B_\infty r^2.
\label{eq:psid_I_sol}
\eeqn
Since $\psi_{d,\I} \gg \psi_\infty$, we may approximate 
the total flux $\psi$ with $\psi_{d,\I}$.
Substituting this into Equation~\eqref{eq:Bz}, 
we finally obtain the vertical field strength deep inside region I, 
\beq
B_{z,\I} \approx \pfrac{\rout}{\rin}^2B_\infty.
\label{eq:Bz_I_asympt}
\eeq
Note that $B_{z,\I}$ is independent of $r$, which is generally the case when diffusion dominates over advection. 

Equation~\eqref{eq:Bz_I_asympt} 
is consistent with the idea that a high conducting disk 
transports all the imposed poloidal flux to region I (see Figure~\ref{fig:split}). 
We denote the flux of the magnetic field threading inside $r < \rin$ by $\Phi$
and the flux of the imposed field threading inside $r < \rout$ by $\Phi'$. 
Since $\Phi \sim \pi \rin^2 B_{z,\I}$ and $\Phi' \sim \pi \rout^2 B_\infty$, 
Equation~\eqref{eq:Bz_I_asympt} implies that $\Phi \approx \Phi'$. 
This understanding is, however, only approximate as $\Phi$ is in fact larger than $\Phi'$ 
by a factor of 2 (see Equation~\eqref{eq:psiin2}).

We also note that the correction factor $\fin$ must be properly taken into account to derive the solution for $B_{z,\I}$ correctly. If one did not include the correction $\fin$ in evaluating the integration in Equation~\eqref{eq:psid_I_sol}, one would obtain 
$\psi_{d,\I} = (1/4)(\rout/\rin)^2B_\infty r^2$, which is two times smaller than the correct expression in Equation~\eqref{eq:psid_I_sol}.

So far we have neglected the current outside region II 
in evaluating the Biot--Savart equation. 
However, as $B_{z,\I}$ and $B_{z,\III}$ can be already known (from Equations~\eqref{eq:BS_0_I} and \eqref{eq:BS_0_III}), $K_{\phi,\I}$ and $K_{\phi,\III}$ can be derived from the steady-state condition (Equation~\eqref{eq:staticcond}), namely, 
\beq
K_{\phi,\I} = \frac{c}{2\pi D_\I} B_{z,\I},
\quad K_{\phi,\III} = \frac{c}{2\pi D_\III} B_{z,\III}.
\label{eq:Kphi_I_III}
\eeq
In fact, these are not necessarily small in magnitude when 
compared to $K_{\phi,\II}$.
However, this does not contradict our assumption because 
their contribution to the induced flux is indeed negligible as long as $D_\II \ll 1$. 

\begin{figure*}
\epsscale{1.15}
\plotone{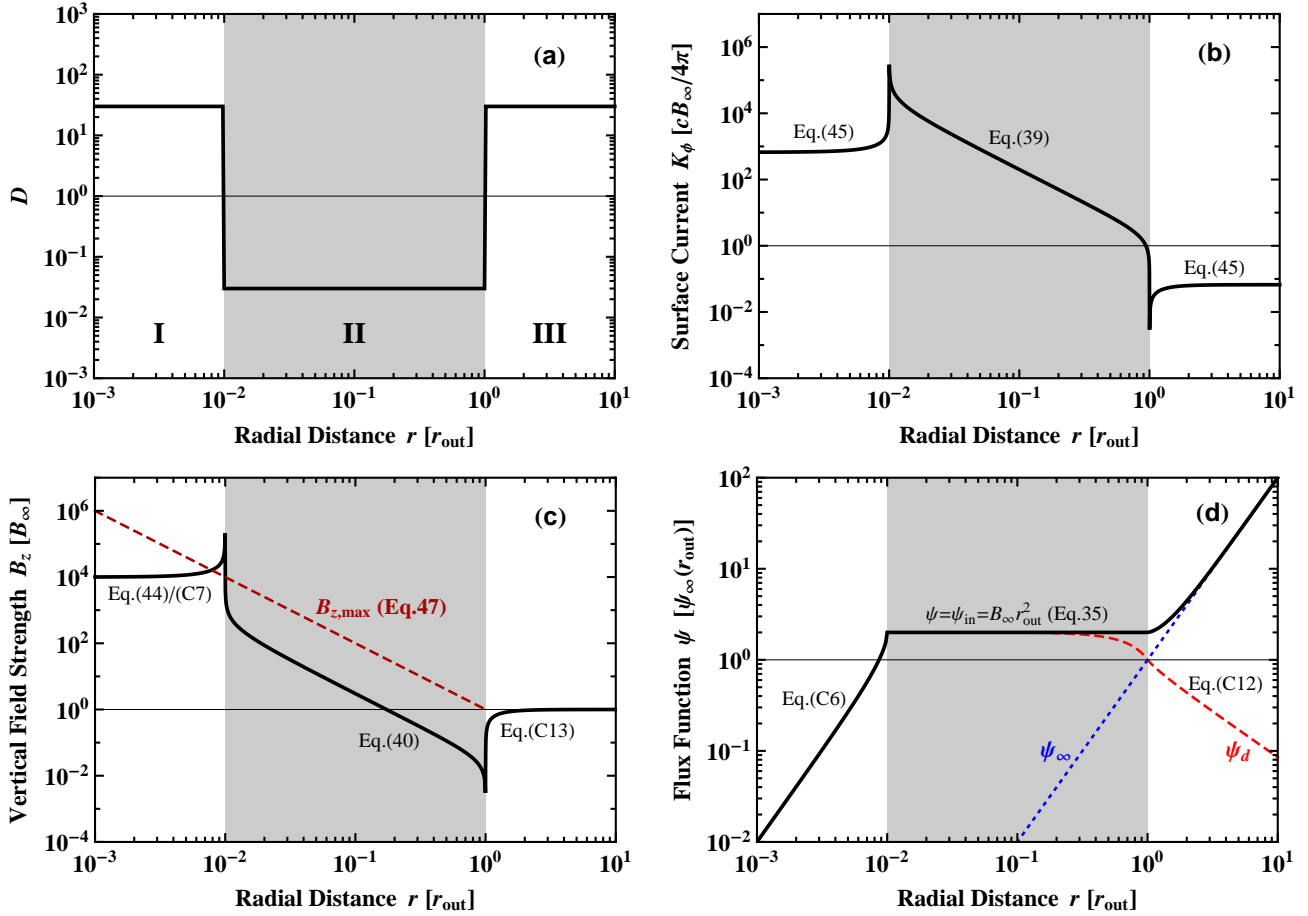}
\caption{Steady-state solution for a piecewise constant $D$.
The inner boundary of region II is set to $\rin = 0.01\rout$. 
Panel (a) displays the assumed radial profile ($D_\I = D_\III =  0.03$, and $D_\II = 30$).
Panels (b), (c), and (d) show 
the surface current density $K_\phi$, 
vertical field strength $B_z$, and flux function $\psi$, respectively.
The flux function is the sum of the contributions from the imposed fields, 
$\psi_\infty$ (dotted curve), and from 
the induced field, $\psi_d$ (dashed curve).
Note that the spikes and valleys in the plots of $B_z$ and $K_\phi$ merely reflect the discontinuities of the assumed $D$ at the boundaries $r = \rin$ and $\rout$; these features do not appear when $D$ is smooth (see Section~\ref{sec:smoothD} and Figure~\ref{fig:comp_smooth}).
}
\label{fig:DKB}
\end{figure*}
As an example, we illustrate in Figure~\ref{fig:DKB} the steady solution for the case of 
$\rin = 0.01\rout$, $D_\I = D_\III = 30$, and $D_\II = 0.03$.
The same set of parameters will be considered in Section~\ref{sec:comp} where we will compare the analytic steady solution with time-dependent solutions from direct numerical integration. We will see that this steady solution is indeed materialized as a result of time evolution. 

\subsection{Maximum Field Strength as a Function of $r$}\label{sec:Bzmax}
An important implication of the steady-state solution is the 
presence of an {\it upper limit} on the vertical field strength $B_z$
attainable in steady states.
To see this, let us consider how $B_z$ at a fixed position $r$ $(< \rout)$ changes with changing the position of the inner boundary $\rin$.
We neglect the correction factors $\fin$ and $\fout$ as they become 
important only when $r$ is very close to $\rin$ or $\rout$ (furthermore, we will see in Section~\ref{sec:smoothD} that the features of these correction factors 
are smeared when $D$ is smooth). Then, from Equations~\eqref{eq:Bz_II_sol} and \eqref{eq:Bz_I_asympt}, we find 
\beq
B_z  = \left\{ \begin{array}{ll}
\pfrac{\rout}{\rin}^2 B_\infty,  & r < \rin ~~ (\textrm{region I}), \\
D(r) \pfrac{\rout}{r}^2 B_\infty, & r > \rin ~~ (\textrm{region II}).
\end{array} \right.
\eeq
Now let us view this as a function of $\rin$ and observe how $B_z$ varies as  
$\rin$ varies from $> r$ to $ < r$.
For $\rin > r$, $B_z$ monotonically increases with decreasing $\rin$ 
and reaches a maximum $(\rout/r)^2 B_\infty$ at $\rin = r$. 
Once $\rin$ falls below $r$, $B_z$ does not exceed this value any more 
since $D < 1$ in region II.
Thus, in the steady states, $B_z$ at a given position is bounded from above by 
\beq
B_{z,{\rm max}}(r) = \pfrac{\rout}{r}^{2}B_\infty.
\label{eq:Bzmax}
\eeq
Note that Equation~\eqref{eq:Bzmax} does not apply to region III ($r>\rout$), 
where $B_z \approx B_\infty$ (see Figure~\ref{fig:DKB}(c)).
The dashed line in Figure~\ref{fig:DKB}(c) indicates Equation~\eqref{eq:Bzmax} for $r<\rout$.

\section{Comparison with Numerical Solutions}\label{sec:comp}
In this section, we test the steady-state solution obtained in Section~\ref{sec:analytic} with numerical calculations of the original evolution equation (Equation~\eqref{eq:psi_avr}). 
We integrate Equations~\eqref{eq:psi_avr} and \eqref{eq:BS} using the finite-volume method described by \citetalias{LPP94a}.
We set $\rin = 0.01\rout$ and take the computational domain
to be $r_{\rm min} \leq r \leq r_{\rm max}$ with $r_{\rm min} = 0.03\rin$ and $r_{\rm max} = 30\rout$. The domain is divided into 400 logarithmically spaced cells. We use a zero-flux boundary condition for the inner boundary (i.e., $\pd\psi/\pd t = 0$ at $r=r_{\rm min}$), while for the outer boundary we allow the magnetic flux to flow across it by assuming that $B_z =B_\infty$ (or equivalently $\pd\psi/\pd r = r B_\infty$) at $r=r_{\rm max}$.

The time-dependent problem requires either $\eta_*$ or $u_*$ to be given in addition to $D$. Here we prescribe $\eta_*$ to be a quadratic function of $r$, 
$\eta_*(r) =r^2/ t_*$, where the constant $t_*$ has the dimension of time.
The radial distribution of $u_*$ is then determined according to the definition of $D$ (Equation~\eqref{eq:D}), i.e., $u_* = -\eta_*/(DH)$.
Thus, the advection speed $|u_*|$ is small in regions I and III 
and is large in region II.
The disk thickness $H$ is taken so that the disk aspect ratio $h \equiv H/r$
is $0.1$ for all $r$.  
All these input parameters are assumed to be independent of time $t$ for simplicity. 
This restriction will be relaxed in \citetalias{TO14}.

The initial condition is specified by the initial distribution of the 
disk-induced flux $\psi_d$. 
Fiducially we will take $\psi_d(t=0) = 0$ (i.e., $B_z = B_\infty$) 
throughout the computational domain, 
but will also consider the case where the innermost region (region I) initially possesses an induced field in excess of the steady-state value. 

In the following subsections, we consider two model functions for $D$ and 
present the results for each model.
 
\subsection{Piecewise Constant $D$}\label{sec:piecewiseD}
\begin{figure}
\epsscale{1.15}
\plotone{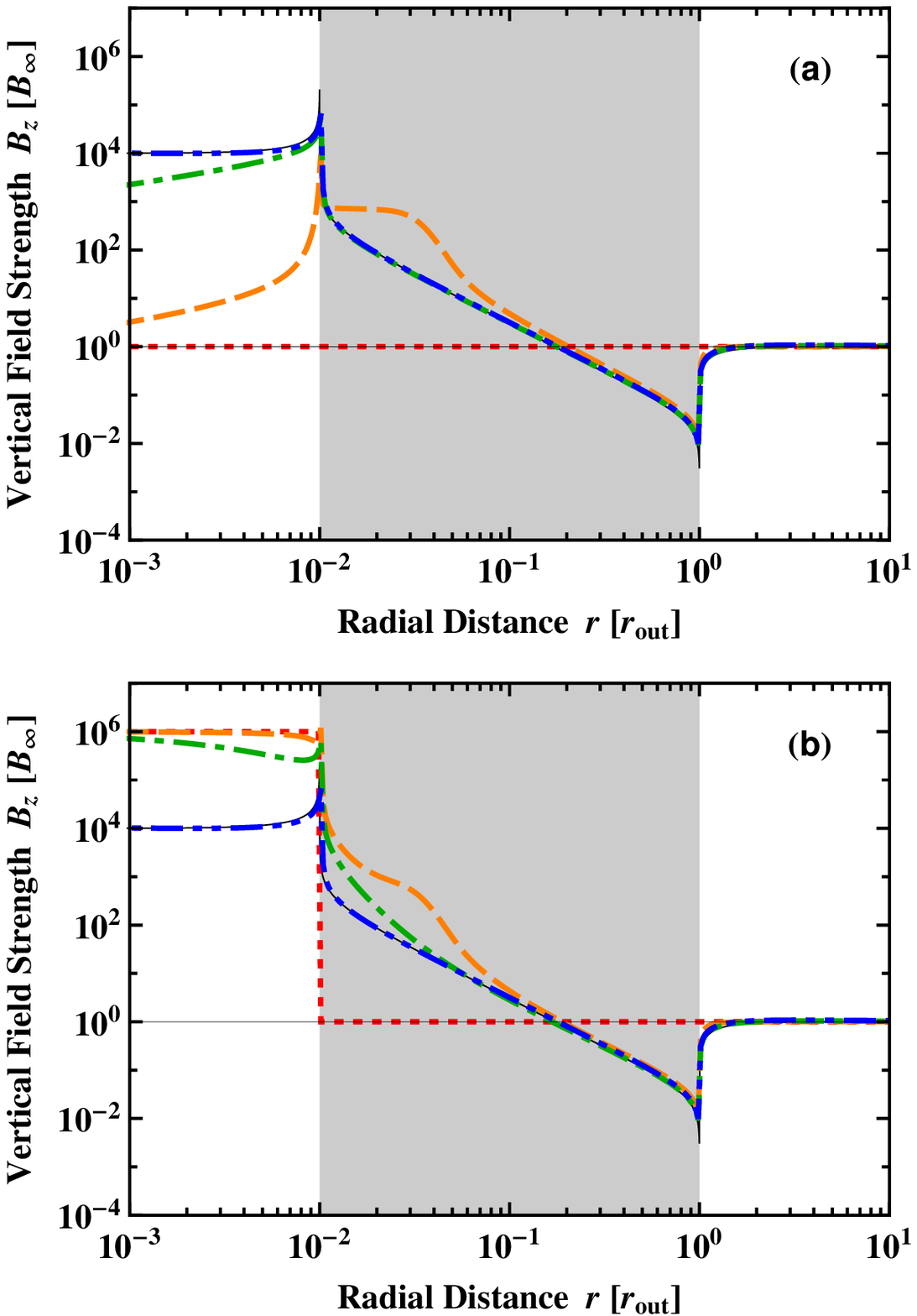}
\caption{Comparison between analytic and time-dependent numerical solutions of Equations~\eqref{eq:psi_avr} and \eqref{eq:BS} for piecewise constant $D$ (Section~\ref{sec:piecewiseD}; see Figure~\ref{fig:DKB}(a) for the radial profile of the assumed $D$).
Panel (a) shows the result for the initial condition $B_z = B_\infty$, 
while in the simulation shown in panel (b) the initial $B_z$ is augmented by factor $10^6$ at $r < \rin$.  
The dotted lines show the initial condition.
The dashed, dot-dashed, and dot-dot-dashed lines are the snapshots of the numerical solutions at times $t =0.01t_*$, $0.1t_*$, and $1.0t_*$, respectively.
The solid lines are the analytic steady-state solution derived in Section~\ref{sec:analytic} (also shown in Figure~\ref{fig:DKB}(b)). Note that the snapshots for $t=1.0t_*$ and analytic solution are almost indistinguishable. 
}
\label{fig:comp_piecewise}
\end{figure}
The first model for $D$ is a piecewise constant function given by
$D_\I = D_\III = 30$ and $D_\II = 0.03$. This has already been used  
in Section~\ref{sec:Bz} for an illustrative purpose. 
The radial profile of $D$ is shown in Figure~\ref{fig:DKB}(a).
For the initial condition, we adopt either $B_z = B_\infty$ for all $r$ (case A) 
or $B_z = 10^6B_\infty$ for $r<\rin$ and $B_z = B_\infty$ for $r>\rin$ (case B).

Figures~\ref{fig:comp_piecewise}(a) and (b) display
the results for the two different initial conditions (cases A and B, respectively).
The dotted curves are the initial profiles of $B_z$, 
and the thin solid curves are the analytic steady solution 
for the assumed $D$ (also shown in Figure~\ref{fig:DKB}(b)). 
For both cases, we find that the time-dependent numerical solution 
approaches the analytic steady solution after a sufficient time. 
The only important difference is that region I gains the flux in case A 
while it loses the initial excess flux in case B. 
Thus, any vertical flux in excess of the steady-state value diffuses away with time. 
It should be noted that details of the time evolution depend on 
the assumption about $\eta_*$ and are therefore not important here.

\subsection{Smooth $D$}\label{sec:smoothD}
\begin{figure}
\epsscale{1.15}
\plotone{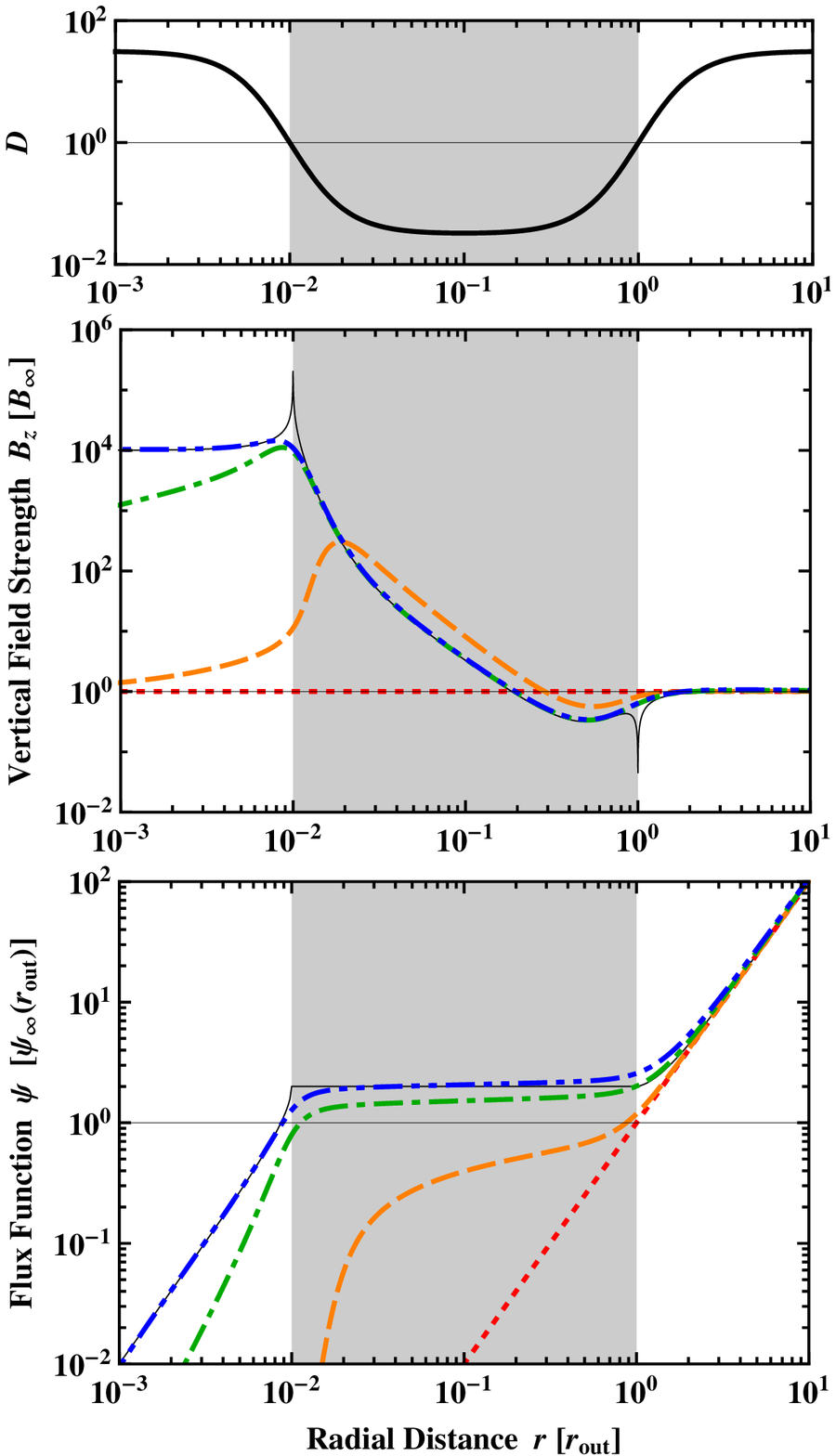}
\caption{Comparison between analytic and time-dependent numerical solutions of Equations~\eqref{eq:psi_avr} and \eqref{eq:BS} for smooth $D$ (Section~\ref{sec:smoothD}). The {top} panel shows the assumed $D$, and the {middle and bottom panels} show the result.  The initial condition is $B_z = B_\infty$ for all $r$ (dotted line). The dashed, dot-dashed, and dot-dot-dashed lines are the snapshots of the numerical calculations at $t =0.01t_*$, $0.1t_*$, and $1.0t_*$, respectively.
}
\label{fig:comp_smooth}
\end{figure}
So far we have assumed that $D$ sharply varies  
at the boundaries of region II. 
We here examine wheter our steady solution is applicable even when $D$ smoothly connects between advective ($D<1$) and diffusive ($D>1$) regions.
We adopt a continuous function $\log D = 1.5\tanh[3\log(r/\rin)]\tanh[3\log(r/\rout)]$.
This smoothly crosses unity at $r = \rin$ and $r=\rout$ 
as depicted in the top panel of Figure~\ref{fig:comp_smooth}.
For this model, we predict the steady-state profile of $B_z$ 
in the same way as we did for piecewise constant $D$, i.e., 
by applying Equations~\eqref{eq:Bz_I_exact}, \eqref{eq:Bz_II_sol}, and
\eqref{eq:Bzd_III_exact} for regions I, II, and III, respectively. 
The predicted profile of $B_z$ is shown in the middle panel 
of Figure~\ref{fig:comp_smooth} with a solid curve. 
Note that the profile of $B_z$ deep inside region II ($\rin \ll r \ll \rout$) 
is no longer given by a power law because of the radial dependence of $D$.

We compare this with the numerical solution of the time-dependent problem 
with the initial condition $B_z = B_\infty$.
We again see that the analytic solution well reproduces the numerical solution at late times. One exception is that the numerical solution has no visible spike or valley in $B_z$ near  $r = \rin$ and $r=\rout$ (see Figure~\ref{fig:comp_piecewise}).
One might expect that the value of $\psi$ in region II should be 
smaller than the prediction from the analytic solution
because the peak of $B_z$ is absent at $r = \rout$. 
However, inspection of $\psi$ shows that its steady-state value agrees with 
the prediction $2\psi_\infty(\rout)$ (Equation~\eqref{eq:psiin}) 
as long as one measures it slightly outside $r = \rout$. 
This suggests that the sharp peak in $B_z$ is not removed but rather smeared out by the smoothed $D$ profile. 
Therefore, we conclude that our analytic steady-state solution is applicable even when $D$ is a smooth function of $r$, except for smeared structures near the positions where $D=1$.

\section{Application to Protoplanetary Disks}\label{sec:application}
\begin{figure}[t]
\epsscale{1.15}
\plotone{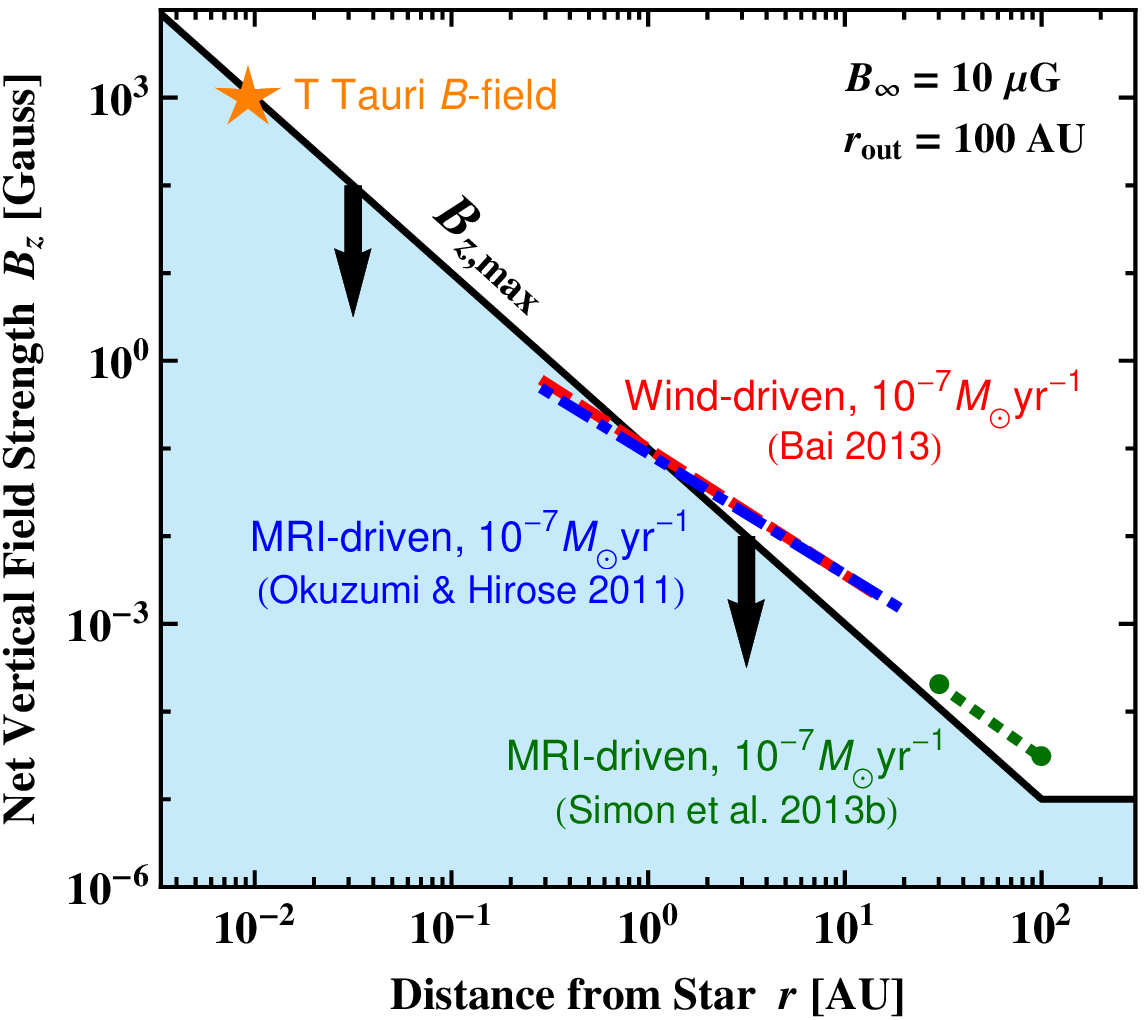}
\caption{Upper limit on the large-scale vertical field strength, 
$B_{z,\rm max}$ (Equation \eqref{eq:Bzmax2}; solid line), 
for $B_\infty = 10~{\rm \mu G}$ and $\rout = 100~\AU$. 
The dot-dashed, dashed, and dotted lines show theoretical predictions 
for the vertical field strength required for disk accretion of 
$\dot{M} = 10^{-7}~M_\odot~{\rm yr}^{-1}$ (see Section~\ref{sec:Mdot} for details). 
The star symbol marks the typical field strength and radius of 
observed T Tauri stars.
}
\label{fig:B_PPD}
\end{figure}

In Section~\ref{sec:Bzmax}, we have derived the maximum strength on 
the vertical field strength in steady states. 
We here apply this result to protoplanetary disk systems 
and discuss its astrophysical implications. 
As a reference value, we take $\rout = 100~\AU$ from 
millimeter observations of T Tauri disks \citep[e.g.,][]{K+02,A+09},
and $B_\infty = 10~{\rm \mu G}$ from Zeeman observations of molecular clouds
\citep[e.g.,][]{TC08,CHT09}.
Then, Equation~\eqref{eq:Bzmax} predicts 
\beq
B_{z,{\rm max}}(r) = 0.1 r_\AU^{-2}\pfrac{\rout}{100~{\rm AU}}^{2}\pfrac{B_\infty}{10~{\rm \mu G}}~{\rm G},
\label{eq:Bzmax2}
\eeq
where $r_\AU = r/(1~\AU)$. 
The solid line in Figure~\ref{fig:B_PPD} shows 
the radial distribution of $B_{z,\rm max}$ 
for the case of $B_\infty = 10~{\rm \mu G}$ and $\rout = 100~\AU$.

{ In protoplanetary disks, the magnetic pressure given by Equation~\eqref{eq:Bzmax2}
is much smaller than the gas pressure unless the disk gas is significantly depleted.
In this sense, the large-scale field itself has no effect on the dynamics of the gas disks near the midplane.
For example, if we take the gas pressure distribution from the minimum-mass solar nebula 
model of \citet{H81} and the magnetic pressure from Equation~\eqref{eq:Bzmax2}, 
then the gas-to-magnetic pressure ratio (or the plasma beta) at the midplane is 
$\sim 10^6$ at $r = 100$ AU and $\sim 10^4$ at $r=0.1$ AU. 
However, MRI and disk winds induced by the large-scale field do 
have a significant effect on disk evolution. We will see this in Section~\ref{sec:Mdot}.} 

\subsection{Upper Limits on the Accretion Rate}\label{sec:Mdot}
The primary purpose of this study is to understand 
the evolution of protoplanetary disks in the framework 
of magnetic accretion scenarios.  
As mentioned earlier, both MRI-driven and wind-driven 
accretion mechanisms predict that the mass accretion rate $\dot{M}$
depends on the strength of the net (large-scale) vertical field. 
In this subsection, we synthesize Equation~\eqref{eq:Bzmax2} and existing 
theoretical predictions for the $\dot{M}$--$B_z$ relation to constrain 
the range of the accretion rate attainable by these accretion mechanisms.

Because protoplanetary disks are poorly ionized, 
MRI and winds in the disks are highly susceptible to non-ideal MHD effects.
Lacking an established accretion model accounting for all the non-ideal effects,
we employ two existing models in parallel.   
We first consider the classical scenario where
only ohmic diffusion is taken into account. 
For this case, \citet{OH11} empirically obtained the $\dot{M}$--$B_z$ relation
for MRI-driven turbulence.
They found that if ohmic diffusion suppresses MRI near the midplane
then the relation is given by\footnote{Equation~\eqref{eq:alpha_OH11} follows from 
{ the relations $\alpha = \alpha_{\rm core} + \alpha_{\rm atm}$, $\alpha_{\rm atm} \gg \alpha_{\rm core}$,
and $\alpha_{\rm atm} \approx 530 \beta_{z0}^{-1}$, which hold when the MRI-dead zone is thick (for details, see Section 5.2 of \citealt{OH11}).}
}
\beq
\alpha \approx 530 \beta_{z0}^{-1}.
\label{eq:alpha_OH11}
\eeq
Here $\alpha$ is the height-averaged turbulent stress normalized by the height-averaged pressure (also known as the Shakura--Sunyaev $\alpha$ parameter), 
and $\beta_{z0}$ is the plasma beta for the net vertical flux measured at the midplane. 
In steady states, $\alpha$ is related to $\dot{M}$ as 
$\dot{M} = {2\pi \alpha \Sigma c_s^2}/{\Omega}$,
where $\Omega$ is the orbital frequency, $\Sigma$ is the gas surface density, 
and $c_s$ is the sound speed (see, e.g., Section~III.A of \citealt*{BH98}; Equation~(19) of \citealt*{S+13a}).
The midplane plasma beta is related to the net vertical field strength $B_z$ as
$\beta_{z0} = {8\pi \Sigma c_s^2}/{\sqrt{2\pi} B_z^2 H_g }$,
where $H_g = c_s/\Omega$ is the pressure scale height of the disk.
If we use these relations, Equation~\eqref{eq:alpha_OH11} translates into  
the $\dot{M}$--$B_z$ in physical units,
\beq
\dot{M} \approx 330\frac{H_g}{\Omega} B_z^2.
\label{eq:Mdot_OH11_0}
\eeq
Note that $\dot{M}$ is {\it independent} of the disk mass 
since $\alpha \propto \dot{M}/\Sigma$ while $\beta_{z0} \propto \Sigma/B_z^2$.
This reflects the fact that the Maxwell stress does not explicitly 
involve the gas density (see \citealt*{W07} and \citealt*{B11a} for a similar expression).  

Let us see how $B_{z,\rm max}$ 
constrains the accretion rate given by Equation~\eqref{eq:Mdot_OH11_0}.
Here we assume a protoplanetary disk around a solar-type star and adopt
$\Omega = 2.0\times 10^{-7}r_\AU^{-3/2}~{\rm s^{-1}}$ and $H_g = 0.033r_\AU^{5/4}~\AU$ following \citet{H81}.
Substituting these into Equation~\eqref{eq:Mdot_OH11_0}, we obtain
\beq
\dot{M} \approx 1.3\times 10^{-7} r_\AU^{11/4}\pfrac{B_z}{0.1~{\rm G}}^2 
M_\odot~{\rm yr}^{-1}.
\label{eq:Mdot_OH11}
\eeq
This gives the MRI-driven accretion rate for given $B_z$.
Applying the constraint $B<B_{z,\rm max}$ to the above relation,
we obtain $\dot{M} \la 1\times 10^{-7} M_\odot~{\rm yr^{-1}}$ 
at $r=1~{\rm AU}$ and 
$\dot{M} \la 0.7\times 10^{-8} M_\odot~{\rm yr^{-1}}$
at $r = 10~\AU$. 
The {dot-dashed} line in Figure~\ref{fig:B_PPD} shows the value of $B_z$ required 
for $\dot{M} = 10^{-7}M_\odot~{\rm yr^{-1}}$ 
predicted by Equation~\eqref{eq:Mdot_OH11}.

A similar constraint is derived from the more recent scenario 
including ambipolar diffusion. 
Recent MHD simulations by \citet{BS13b} and \citet{B13} 
have suggested that accretion at $0.3~\AU \la r \la 15~\AU$ 
is mainly driven by magnetocentrifugal winds because 
ambipolar diffusion completely suppressed MRI there. 
The relation between the wind-driven $\dot{M}$ and $B_z$ has been provided by \citet{B13}, which reads\footnote{\citet{B14} reports that Equation~(11) of \citet{B13} was provided with error.
Here we use the correct result given by \citet[][his Equation (33)]{B14}.}  
\beq
\dot{M} = 0.47 \times 10^{-8} r_\AU^{1.90} \pfrac{B_z}{10~{\rm mG}}^{1.32} M_\odot~{\rm yr^{-1}}.
\label{eq:Mdot_B13}
\eeq 
Note that $\dot{M}$ is again independent of $\Sigma$
 (see \citealt*{B13} for more discussions).
If we apply $B<B_{z,\rm max}$ to Equation~\eqref{eq:Mdot_B13}, we obtain 
{ $\dot{M} \la 1\times 10^{-7} M_\odot~{\rm yr^{-1}}$}
at $r=1~{\rm AU}$ and $\dot{M} \la 2\times 10^{-8} M_\odot~{\rm yr^{-1}}$ at $r = 10~\AU$. 
The dashed line shows the value of $B_z$ required for $\dot{M} = 10^{-7}M_\odot~{\rm yr^{-1}}$ predicted by Equation~\eqref{eq:Mdot_B13}.

Farther out in the disks, MRI operates even in the presence of ambipolar diffusion. 
MHD simulations by \citet{S+13b} show 
that $\dot{M}$ increases with $B_z$
and reaches $10^{-7}M_\odot~{\rm yr^{-1}}$ at $B_z \sim 200~{\rm \mu G}$ 
for $r=30~\AU$, and at $ 30~{\rm \mu G}$ for $r=100~\AU$.
These values are marked in  Figure~\ref{fig:B_PPD} by the filled circles connected by the dotted line. We find that the required field strengths 
are close to the upper limit $B_{z,\rm max}$, which suggests 
that $\dot{M} \la 10^{-7}M_\odot~{\rm yr^{-1}}$ for these regions.

Taken together, all these models suggest that  
the upper limit on the magnetically driven accretion rate is 
$\sim 10^{-7}M_\odot~{\rm yr^{-1}}$. 
This is consistent with upper limits suggested by observations \citep{H+98,C+04,S-A+05}.

\subsection{Possible Relevance to the ``Magnetic Flux Problem'' in Star Formation}

Another interesting application of Equation~\eqref{eq:Bzmax2} is
to the so-called magnetic flux problem in star formation \citep{N84}.
Young stars are known to have a magnetic field of 
typical strength $\sim {\rm kG}$ \citep[e.g.,][]{J07}.
Star-forming dense cores of molecular clouds also have a magnetic field \citep{TC08,CHT09}, but the magnetic flux of a single core is
about four orders of magnitude larger than that of a single young star. 
Therefore, the magnetic flux that was inherited from the parent cloud 
must have dissipated, at least to a $\la {\rm kG}$ level, 
by the time the star became visible.
ohmic dissipation during star formation is 
one plausible solution to this problem \citep[e.g.,][]{NNU02,S+06,MIM07,DBK12}. Other possibilities include flux destruction after the external field detaches from the star \citep{B12}.

Here we explore the possibility that the excess flux is lost through 
transport between the star and surrounding protoplanetary disk.
Let us suppose that a protoplanetary disk extends to a central star
and hence Equation \eqref{eq:Bzmax2} is applicable down to the interface 
between the disk and star.
Then, Equation \eqref{eq:Bzmax2} suggests that 
$B_{z,\rm max} \approx 1~{\rm kG}$ at protostellar radius 
$\approx 2R_\odot \approx 10^{-2}~{\rm AU}$ for the selected parameters 
$B_\infty = 10~{\rm \mu G}$ and $\rout = 100~\AU$ 
(see the star symbol in Figure~\ref{fig:B_PPD}).  
Interestingly, the predicted $B_{z,\rm  max}$ is in good agreement 
with the observed stellar field strengths.
Any excess flux above this level will be lost after a sufficient time
(as demonstrated in Figure~\ref{fig:comp_piecewise}(b)) 
if the flux transport is allowed across the star--disk interface
(for example, convection inside the star might effectively diffuse the flux toward the stellar surface). 
If this mechanism is viable, then this might partly resolve 
the long-standing problem of star formation theory.
Further examination of this possibility will require a more detailed model 
for the star--disk interface (e.g., a magnetosphere and a jet will need to be taken into account), 
which is beyond the scope of this paper.

\section{Summary and Discussion}\label{sec:summary}
We have studied how a large-scale poloidal field 
is transported in a highly conducting accretion disk, 
primarily aiming at understanding the viability of magnetically driven 
accretion mechanisms.
We have employed the kinematic mean-field model 
for a large-scale poloidal field originally developed by \citetalias{LPP94a}.  
As a first step, we have focused on steady states where inward 
advection of a poloidal field balances with outward diffusion.
In a companion paper \citep{TO14}, we extend our analysis 
to time-dependent problems and study how poloidal fields are transported 
in evolving accretion disks.  

We have analytically derived the steady-state distribution of 
a poloidal field for conducting accretion disks.
The most important finding from this solution is that 
there is an upper limit on the large-scale vertical field
strength attainable in steady states (Equation~\eqref{eq:Bzmax}).
The upper limit is given by a function of the distance from the central star ($r$), 
the poloidal strength at infinity ($B_\infty$), and the outer radius of the highly conductive region ($\rout$). 
For $\rout = 100~\AU$ and $B_\infty = 10~{\rm \mu G}$, 
the maximum vertical field strength is about $0.1~{\rm G}$ at $r= 1~\AU$, 
and about $1~{\rm mG}$ at $r=10~\AU$ (Equation~\eqref{eq:Bzmax2}; Figure~\ref{fig:B_PPD}). 
Any poloidal field of a strength above the limit will eventually 
diffuse away outward on the diffusion timescale. 
We have demonstrated this with time-dependent numerical calculations 
of the mean-field equations. 

We have applied this upper limit to a large-scale poloidal field 
threading a protoplanetary disk around a young star. 
We have adopted three different theoretical models for magnetically driven accretion \citep{OH11,B13,S+13b} to translate the large-scale vertical field strength into the accretion rate.
All three models suggest a maximum steady-state 
accretion rate of $\sim 10^{-7}~M_\odot~{\rm yr}^{-1}$.
This is in agreement with observations showing that 
$\dot{M} \sim 10^{-9}$--$10^{-7}~M_\odot~{\rm yr}^{-1}$ for T Tauri stars.

We have also applied the upper limit to the flux of young stars 
assuming that surrounding disks extend to the stellar surface.
We find that the maximum field strength is $\sim 1~{\rm kG}$ 
at the surface of the central star.
This implies that any excess stellar poloidal field of strength $\ga {\rm kG}$ 
can be lost via outward diffusion to the circumstellar disks.
That mechanism might explain why observed young stars
have a significantly small magnetic flux compared to molecular cloud cores.
Further examination of this possibility will await a detailed model for the gas dynamics 
on the star--disk boundary.

As already pointed out by \citet{OL01},
the details of the steady-state solution depend on the vertical structure 
of the disk through the dimensionless quantity $D$ (Equation~\eqref{eq:D}).
Although recent studies suggest $D<1$ (see the introduction),  
the precise value of $D$ is still uncertain for real accretion disks. This is particularly true 
for protoplanetary disks, where non-ideal MHD effects are likely to affect the value of it. 
In this study, we have assumed $D \ll 1$ for the bulk of the disks motivated by the fact that otherwise the vertical magnetic flux would become too small to affect disk evolution \citep{BS13b,S+13a}.  
This assumption needs to be justified with MHD simulations that take into account 
the ionization structure of the disks in the vertical direction. 
Very recently, \citet{BS13b} and \cite{B13} have conducted such simulations, and the results show 
that the poloidal field is significantly bent on the disk surface, i.e., 
$B_{rs} > B_z$ (see Figures 10 and 11 of \citealt*{BS13b}). 
This, together with Equation~\eqref{eq:Bratio_static} in this paper, suggests
that $D < 1$ is possible in protoplanetary disks even in the presence of non-ideal MHD effects.

The present study has employed a number of simplifications,
and there are at least three important caveats that should be mentioned. 
First, all important parameters have been treated as time independent, 
but this assumption breaks down when the gas disk evolves faster than the poloidal field. 
If this is the case, the poloidal-field distribution largely depends 
on the initial condition, as we will discuss in a companion paper \citep{TO14}. 
Second, we have assumed that the toroidal current vanishes 
in the exterior of the disk and approximated the exterior field as a potential field.
However, it is not obvious how this approximation applies when a magnetocentrifugal wind is present.
This issue was already noted and discussed in detail by \citet{O97}. 
He showed that the Biot--Savart equation precisely describes the exterior 
field as long as the field inside the Alfv\'{e}n surface of the wind is considered. 
However, the asymptotic field $B_\infty$ must then be regarded as 
including the contribution from the toroidal currents beyond the Alfv\'{e}n surface. 
This contribution was neglected in this study and therefore will need to be quantified in future work. 
 Finally, it is not always true that a steady state is reached as a result of disk--field evolution. 
For example, some models predict that wind-driven accretion is unstable \citep{LPP94b,CS02,M12}.

\acknowledgments
We thank Takeru Suzuki, Shu-ichiro Inutsuka, and Hidekazu Tanaka for inspiring discussion.
We also thank the anonymous referee for comments that clarified the paper. 
This work was supported by Grant-in-Aid for Research Activity Start-up (\#25887023, {24840037}) from JSPS, and by Grants-in-Aid for Scientific Research (\#20540232, 23103005) from MEXT. 
\\[6mm]
{{\it Note added in proof.} Recently, \citet{GO14} have independently studied  
the radial transport of large-scale poloidal fields in protoplanetary disks. 
Their mean-field approach is essentially the same as ours, although they also consider cases 
where the transport coefficients ($u_*$ and $\eta_*$) depend on the vertical field strength. 
For the cases where the transport coefficients are independent of the field strength, 
their results are consistent with ours; in particular, we have numerically verified that their self-similar steady solution 
(their Equation~(31)) is precisely equivalent to ours (our Equation~(A7)). 
 }
\appendix

\section{Asymptotic Solution in region II for Arbitrary Constant $D_\II$}\label{sec:constantD}
In this appendix, we consider cases where $D_\II$ is constant but not necessarily small. For these cases, Equations~\eqref{eq:BS} and \eqref{eq:staticcond} become self-similar with respect to $r$ in the limit of $\rin \ll r \ll \rout$ and $\psi_\infty \ll \psi_d$, and therefore the solution must be written as a power law. Specifically, if we suppose $K_\phi \propto r^{-q}$ with $q$ being a constant, then a dimensional analysis shows $B_z \propto r^{-q}$ and $\psi \propto r^{2-q}$. As we will see below, the exponent $q$ is determined as a function of $D_\II$.

Below we omit the subscript ``II'' for the sake of clarity. 
In the limit of $\rin \ll r \ll \rout$ and $\psi_\infty \ll \psi_d$, Equation~\eqref{eq:BS} reduces to 
\beq
\psi = \frac{4}{c}\int_0^\infty R(r_<,r_>)K_\phi(r') dr'.
\label{eq:BS_hom}
\eeq
Note that the perturbative approach employed in Section~\ref{sec:perturb} is not used here as it only applies to $D\ll 1$. 
Let us assume 
\beq
K_\phi(r)= A r^{-q},
\label{eq:Kphi_power}
\eeq
where $A$ and $q$ are constants. 
Substituting Equation~\eqref{eq:Kphi_power} 
into Equation~\eqref{eq:BS_hom}, we obtain
\beq
\psi
= \frac{2\pi C(q)}{c} A r^{2-q}, 
\label{eq:psi_power}
\eeq
where 
\beqn
C(q) 
&\equiv& \frac{2}{\pi} \int_0^{1} 
\left[\EllipticK(x) - \EllipticE(x)\right](x^{-q} + x^{q-3})dx
\label{eq:Cq_def}
\eeqn
is a numerical coefficient that depends on $q$.
For $0<q<3$, the integration on the right-hand side 
of Equation~\eqref{eq:Cq_def} converges, resulting in
\beq
C(q) = \frac{2}{3\pi} \left[ 
{_3}{F}_2\left(1,1,\tfrac{4-q}{2};\tfrac{3}{2},\tfrac{5}{2};1\right)+
 {_3}{F}_2\left(1,1,\tfrac{q+1}{2};\tfrac{3}{2},\tfrac{5}{2};1\right) 
\right],
\label{eq:Cq}
\eeq
where ${_3}{F}_2$ is a generalized hypergeometric function.
It is useful to note that $C(1) = C(2) = 1$.

\begin{figure}[t]
\epsscale{1.15}
\plotone{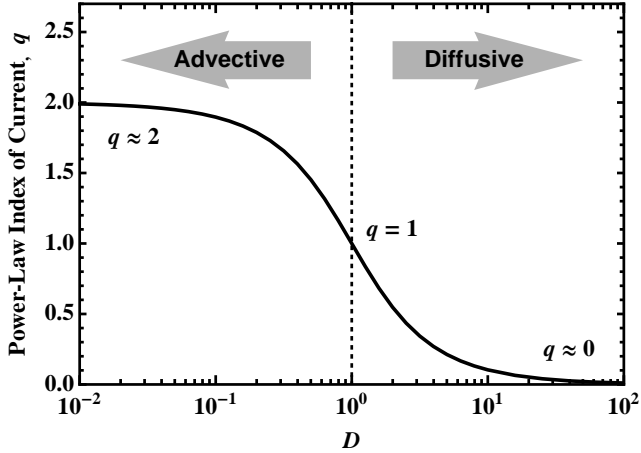}
\caption{Power-law index $q$ for radially constant $D$.}
\label{fig:q}
\end{figure}

The induced vertical field  is 
\beqn
B_z &=& 
\frac{1}{r}\frac{d\psi}{dr} = 
\frac{2\pi(2-q)C(q)}{c} A r^{-q}
\nonumber \\
&=& 
\frac{2\pi(2-q)C(q)}{c} K_\phi.
\label{eq:Bzd_q}
\eeqn
Comparing this with Equation~\eqref{eq:staticcond}, 
we obtain the equation for$q$,
\beq
(2-q)C(q) = D.
\label{eq:q}
\eeq
Figure~\ref{fig:q} plots the solution of Equation~\eqref{eq:q}
as a function of $D$.
We see that $q$ monotonically increases 
with decreasing $D$, 
meaning that the radial profiles of $B_z$ and $K_\phi$ 
become steeper as inward advection becomes more effective.
The asymptotic values are $q \to 2$ for $D \to 0$ and 
$q \to 0$ for $D \to \infty$.
The solution for $D \to \infty$ just shows that 
the vertical field strength becomes uniform in $r$ 
in the limit where diffusion dominates over advection. 

\section{Inverting Biot--Savart's Integral Equation}\label{sec:BS_append}
In this appendix, we present an analytic method to invert  
Biot--Savart's  equation.
In an integral form, the Biot--Savart's law can be written as 
\citep[see, e.g.,][]{J98}
\beq
\frac{\psi_d(r)}{r} = \frac{1}{c}\int_{\rin}^{\rout} r' dr'\int_0^{2\pi} d\phi'
\frac{K_\phi(r') \cos\phi'}{(r^2+r'^2 - 2 r r'\cos\phi')^{1/2}}
\label{eq:BS_appendix}
\eeq
(as shown below, this is equivalent to Equation~\eqref{eq:BS} in the main text).
We seek a $K_\phi$ that gives a flux function of the form
$\psi_d \propto  {\rm constant} -r^2$ or $\psi_d \propto {\rm constant}$.

To begin with, we perform azimuthal integration in these equations.
The integration can be written in terms of the Laplace coefficient known in
celestial mechanics \citep[see, e.g.,][]{MD99}.
The series expansion of the Laplace coefficient gives 
\beq
\int_0^{2\pi}\frac{\cos \phi' \, d\phi'}{(r^2 + r'^2 - 2rr' \cos\phi')^{1/2}}
= \frac{\pi}{r_>}\sum_{n=0}^\infty  c_n \pfrac{r_<}{r_>}^{2n+1},
\label{eq:azimuth_expand}
\eeq
\beq
c_n = \frac{ (\frac{1}{2})_n(\frac{3}{2})_n}{(1)_n(2)_n},
\eeq
where 
$(a)_n = a(a+1)\cdots(a+n-1)$ is the Pochhammer symbol.
One can show, by using Equation~\eqref{eq:azimuth_expand} 
together with the relations $\sum_{n=0}^\infty  c_n \alpha^{2n+1} = (4/\pi\alpha)[\EllipticK(\alpha)-\EllipticE(\alpha)]$ and $rr'/r_< = r_>$,
that Equation~\eqref{eq:BS_appendix} is equivalent to Equation~\eqref{eq:BS}.
Putting Equation~\eqref{eq:azimuth_expand} into Equation~\eqref{eq:BS_appendix}, we obtain Biot--Savart's equation after azimuthal integration,
\beq
\psi_d = \frac{\pi}{c} \sum_{n=0}^\infty c_n
\left[  \frac{1}{r^{2n+1}} \int_{\rin}^r r'^{2n+2} K_\phi dr' 
+ r^{2(n+1)} \int_{r}^{\rout} \frac{K_\phi dr'}{r'^{2n+1}} 
\right],
\label{eq:psid_expand0}
\eeq
for $\rin < r < \rout$. 

\subsection{Analogy with Inverse Problems for Gravitational Fields}
Before we proceed, it is useful to see how our problem resembles  
inverse problems of axisymmetric {\it scalar} fields. 
As pointed out by \citetalias{LPP94a}, 
solving  Biot--Savart's equation for $K_\phi$ 
is analogous to finding the surface density distribution of 
a disk for a given gravitational potential distribution.
For example, for an axisymmetric disk with surface density profile $\Sigma(r)$, 
the gravitational potential $\Phi(r)$ is given by
\beq
\Phi(r) = -G \int_{\rin}^{\rout} r' dr'\int_0^{2\pi} d\phi'
\frac{\Sigma (r')}{(r^2+r'^2 - 2 r r'\cos\phi')^{1/2}},
\label{eq:grav}
\eeq
where $G$ is the gravitational constant.
Equation~\eqref{eq:grav} is identical to Equation~\eqref{eq:BS_appendix}
if we replace $G\Sigma(r') \to K_\phi(r') \cos\phi'$
and $\Phi \to -c\psi_d/r$.
A similar equation can be found in the problem of 
finding the pressure distribution $P(r)$ on 
the interface of two elastic spheres for given surface displacement distribution
$u_z(r)$, in which case $G\Sigma \to P$ 
and $\Phi  \to -\pi E^* u_z$,
where $E^*$ is the reduced Young modulus \citep{J87,LL86}.

It is known that Equation~\eqref{eq:grav} has two types of exact solutions
in the limit of $\rin \to 0$.
The first type is 
\beq
\Sigma(r) = \left[1-\Bigl(\frac{r}{\rout}\Bigr)^2\right]^{1/2}\Sigma_c,
\label{eq:Sigma_H}
\eeq
where $\Sigma_c$ is a constant. 
Equation~\eqref{eq:Sigma_H} gives a harmonic 
gravitational potential
\beq
\Phi(r) = -\frac{\pi^2 G\Sigma_c \rout}{2}
\left[1 - \frac{1}{2}\pfrac{r}{\rout}^2 \right].
\label{eq:Phi_H}
\eeq
The second type is 
\beq
\Sigma(r) = \left[1-\Bigl(\frac{r}{\rout}\Bigr)^2\right]^{-1/2}\Sigma_c,
\label{eq:Sigma_B}
\eeq
where $\Sigma_c$ is a constant. 
Equation~\eqref{eq:Sigma_B} gives a constant potential 
\beq
\Phi = - \pi^2 G\Sigma_c \rout.
\label{eq:Phi_B}
\eeq
In contact mechanics ($\Sigma \to P$ and $\Phi/G  \to -\pi E^* u_z$), 
the first and second types 
are known as Hertz's and Boussinesq's solutions, respectively \citep{J87,LL86}. 
These solutions are useful when we guess the functional form of $K_\phi$ 
for given $\psi_d$.

\subsection{Solution Near the Outer Boundary}\label{sec:out}
We first solve Equation~\eqref{eq:psid_expand0} 
for $\psi_d =\psiin - B_\infty r^2/2$ (Equation~\eqref{eq:psid_II}). 
In this subsection, we focus on how the solution behaves near the outer boundary ($r \sim \rout$), and for this reason we will take $\rin$ to be zero.  
As described in the main text, the asymptotic behavior of the solution $r \ll \rout$ shows that the solution must be of the form $K_\phi(r') = (c\psiin/2\pi r'^2)\fout(r')$ (Equation~\eqref{eq:Kphi_hom_sol}), 
where $\fout$ satisfies $\fout \to 1$ as $r'/\rout \to 0$.

To carry out the integration in Equation~\eqref{eq:psid_expand0}, 
we expand $\fout$ in powers  of $r'/\rout$, 
\beq
\fout(r') = 1 + \sum_{m=0}^\infty a_{2m+1} \pfrac{r'}{\rout}^{2m+1},
\label{eq:Kphi_expand}
\eeq
where $a_{2m+1}$ $(m=0,1,2,\dots)$ are constants, and we have used that $\fout \to 1$ as $r'/\rout \to 0$.
We have assumed that $K_\phi$ does not involve terms $r'^{2m}$ ($m=1,2,\dots$) because such terms would yield unwanted logarithmic terms in the resultant $\psi_d(r)$.

With Equation~\eqref{eq:Kphi_expand},  
the integration in Equation~\eqref{eq:psid_expand0} can be analytically performed.
The result reads 
\beqn
\psi_d &=& \frac{\psiin}{2}
\sum_{n=0}^\infty c_n\left[\frac{1}{2(n+1)}+\frac{1}{2n+1}\right]
\nonumber \\ &&
+ \frac{\psiin}{2}\sum_{m=0}^\infty a_{2m+1} \pfrac{r}{\rout}^{2m+1}
\nonumber \\
&&\quad\times 
\sum_{n=0}^\infty c_n\left[ \frac{1}{2(n+m+1)}+\frac{1}{2(n-m)+1}\right]
\nonumber \\
&&+ \frac{\psiin}{2}\sum_{n=0}^\infty c_n \pfrac{r}{\rout}^{2(n+1)} 
\left[ -\frac{1}{2(n+1)} +\sum_{m=0}^\infty \frac{a_{2m+1}}{2(m-n)-1}
\right].
\nonumber \\
\label{eq:psid_expand}
\eeqn
This result can be simplified if we use the relation
\beqn
&&\sum_{n=0}^\infty c_n\left[ \frac{1}{2(n+m+1)}+\frac{1}{2(n-m)+1}\right]
\nonumber \\
&&= \frac{{_3}{F}_2\left(\tfrac{1}{2},\tfrac{3}{2},m+1; 2, m+2; 1\right)}{2(m+1)}
+\frac{{_3}{F}_2\left(\tfrac{1}{2},\tfrac{3}{2},\tfrac{1-2m}{2};2,\tfrac{3-2m}{2};1\right)}{1-2m}
\nonumber \\
&&= \left\{ \begin{array}{ll}
2, & \quad m=0, \\
0, & \quad m= 1,2, \dots 
\end{array} \right. 
\label{eq:sum2}
\eeqn
(one can verify this with  {\it Mathematica}).
With this relation, we obtain
\beqn
\psi_d &=&
\psiin +  \frac{\psiin a_1  r}{\rout} 
\nonumber \\
&&
+ \frac{\psiin}{2}\sum_{n=0}^\infty c_n \pfrac{r}{\rout}^{2(n+1)} 
\left[ -\frac{1}{2(n+1)} +\sum_{m=0}^\infty \frac{a_{2m+1}}{2(m-n)-1}
\right].
\nonumber \\
\label{eq:psid_comp} 
\eeqn
Comparing this with $\psi_d =\psiin - B_\infty r^2/2$, 
we find that the second term of Equation~\eqref{eq:psid_comp}
must vanish, and that the infinite sum in the third term must leave a term proportional to $r^2$.
Hence, the coefficients $a_1$, $a_3$, $a_5$, $\dots$ must satisfy the relations
\beq
a_1 = 0,
\label{eq:am_cond1}
\eeq
\beq
\sum_{m=1}^\infty \frac{a_{2m+1}}{2(m-n)-1} = \frac{1}{2(n+1)} 
\quad n = 1, 2, \dots .
\label{eq:am_cond}
\eeq

Unfortunately, we could not find the set of $a_{2m+1}$ 
that exactly satisfies Equation~\eqref{eq:am_cond}.
However, an approximate but fully accurate solution can be 
obtained if we note that the flux function under consideration is 
of the same functional form 
as the harmonic gravitational potential (Equation~\eqref{eq:Phi_H}).
By analogy with Equation~\eqref{eq:Sigma_H}, 
let us consider the following ansatz for $\fout$, 
\beq
\fout(r') = 1 + \frac{r'}{\rout} \left\{
\left[1-\pfrac{r'}{\rout}^2 \right]^{\gamma} -1 \right\},
\label{eq:K_ansatz_out}
\eeq
where $\gamma$ is a fitting parameter to be determined below.
We may assume  $\gamma > -1$ since otherwise 
the radially integrated current $2\pi \int_0^{\rout} K_\phi(r')r' dr'$ would diverge.
The ansatz is equivalent to Equation~\eqref{eq:Kphi_expand} with $a_1 = 0$ and 
\beq
a_{2m+1} =  \frac{(-1)^{m}}{m!}\frac{\Gamma(\gamma+1)}
{\Gamma(\gamma+1-m)} , \quad m = 1, 2, \dots .
\label{eq:a_2m+1}
\eeq
This ansatz is useful because the infinite sum $\sum_{m=1}^\infty \frac{a_{2m+1}}{2(m-n)-1}$  has a closed expression 
\beq
\sum_{m=1}^\infty \frac{a_{2m+1}}{2(m-n)-1} 
=
\frac{1}{1+2n}  
+ \frac{\Gamma(-\frac{1}{2}-n)\Gamma(1+\gamma)}
{2\Gamma(\frac{1}{2}-n+\gamma)}. 
\label{eq:am_ansatz}
\eeq
By a least-squares method, we find that the right-hand side of Equation~\eqref{eq:am_ansatz} best approximates that of Equation~\eqref{eq:am_cond} when $\gamma = 0.43$ (Figure~\ref{fig:fit}).

In the main text we have introduced $\psiin$ as an undetermined quantity. 
As we will see below, this quantity is determined as an {\it eigenvalue} of the inversion problem under consideration.  
Substitution of Equations~\eqref{eq:am_cond1} and \eqref{eq:am_cond} into Equation~\eqref{eq:psid_comp} shows that $\psi_d$ must be of the form 
\beq
\psi_d = \psiin
+ \frac{\psiin}{2} 
\left(-\frac{1}{2} +\sum_{m=0}^\infty \frac{a_{2m+1}}{2m-1} \right)
\pfrac{r}{\rout}^{2},
\label{eq:psi_out_sol} 
\eeq
where we have used that $c_0 = 1$.
The infinite sum remaining in the above equation can be evaluated as
$\sum_{m=1}^\infty \frac{a_{2m+1}}{2m-1} \approx -0.502 \approx -1/2$ 
if we use Equation~\eqref{eq:am_ansatz}
with the best-fit $\gamma = 0.43$ (see also Figure~\ref{fig:fit}).
Thus, we find
\beq
\psi_d = \psiin \left[ 1 - \frac{1}{2} \pfrac{r}{\rout}^2 \right]. 
\eeq
Imposing that this is equal to $\psiin - B_\infty r^2/2$, 
we obtain $\psiin = B_\infty \rrout$.

\begin{figure}[t]
\epsscale{1.15}
\plotone{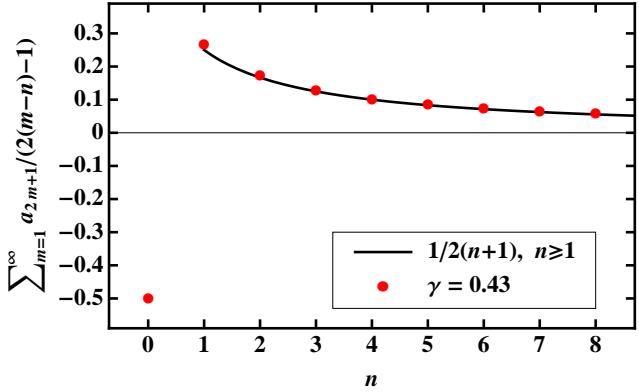}
\caption{
Values of $\sum_{m=1}^\infty \frac{a_{2m+1}}{2(m-n)-1}$
($n=0,1,2,\dots$) for $\gamma = 0.43$ (circles) 
from Equation~\eqref{eq:am_ansatz}.
The curve indicates the right-hand side of Equation~\eqref{eq:am_cond}.
}
\label{fig:fit}
\end{figure}

\subsection{Solution Near the Inner Boundary}\label{sec:in}
Next we show how $K_\phi$ behaves near the inner boundary $r \sim \rin$.
We take $\rout$ to be infinitely large and approximate $\psi_d$ with a constant $\psiin$. As we did for the outer boundary, we assume the solution of the form 
$K_{\phi}(r') = (c\psiin/2\pi r'^2) \fin(r')$, where $\fin$ approaches unity as $r'/\rin \to \infty$. 

Let us expand $\fout$ as
\beq
\fin (r') = 1+ \sum_{m=1}^\infty b_{2m} \pfrac{r'}{\rin}^{-2m},
\label{eq:Kphi_expand_2}
\eeq
where $b_{2m}$ $(m = 1,2,\dots)$ are constants.
We have assumed that $\fin$ is an even function of $r'$
because otherwise the resulting $\psi_d(r)$ would involve unwanted logarithmic terms.
Substituting Equation~\eqref{eq:Kphi_expand_2} 
into Equation~\eqref{eq:psid_expand0}, performing integration,
and using Equation~\eqref{eq:sum2},
we obtain
\beq
\psi_{d}=
 \psiin + \frac{\psiin}{2}\sum_{n=0}^\infty c_n \pfrac{\rin}{r}^{2n+1}  \left[ -\frac{1}{2n+1} + 
\sum_{m=1}^\infty \frac{b_{2m}}{2(m-n)-1} 
\right].
\label{eq:psid_expand_2}
\eeq
Since this must be equal to $\psiin$, 
$b_{2m}$ ($m=1,2\dots$) must satisfy the relation
\beq
\sum_{m=1}^\infty \frac{b_{2m}}{2(m-n)-1}= \frac{1}{2n+1},  
\quad n = 0,1,\dots .
\label{eq:bm_cond}
\eeq

Now let us consider the ansatz
\beq
\fin(r) = 
\left[ 1 - \Bigl(\frac{r'}{\rin}\Bigr)^{-2} \right]^{-\kappa},
\label{eq:K_ansatz_in}
\eeq
where $\kappa$ is a constant (see Equation~\eqref{eq:Sigma_B}).
We require $\kappa < 1$ in order for 
the radially integrated surface current to be finite.
Equation~\eqref{eq:K_ansatz_in} is equivalent to 
Equation~\eqref{eq:Kphi_expand_2} with 
\beq
b_{2m} = \frac{(-1)^{m}}{m!}\frac{\Gamma(1-\kappa)}
{\Gamma(1-m-\kappa)}, \quad m = 1,2,\dots.
\label{eq:b_2m}
\eeq
It can be shown that $b_{2m}$ given by Equation~\eqref{eq:b_2m} satisfies
\beq
\sum_{m=1}^\infty \frac{b_{2m}}{2(m-n)-1} = 
 \frac{1}{2n+1} + 
\frac{\Gamma(-\frac{1}{2}-n)\Gamma(\kappa-1)}
{2\Gamma(\frac{1}{2}-n-\kappa)}.
\label{eq:bm_ansatz}
\eeq
The poles of $\Gamma(\frac{1}{2}-n-\kappa)$
suggest that Equation~\eqref{eq:bm_cond} holds for all $n (= 0,1,2,\dots)$ 
when $\kappa = 1/2, 3/2, 5/2, \dots$. 
Among them, only $\kappa = 1/2$ is the physical solution
since the others do not satisfy $\kappa  < 1$.
Thus, we find that $\fin$ is exactly given by Equation~\eqref{eq:K_ansatz_in} with $\kappa = 1/2$.

\section{Disk-Induced Flux in Regions I and III}\label{sec:C}
Here we derive full analytic expressions for the disk-induced flux $\psi_d$ 
and corresponding field strength $B_{z,d} $ in regions I and III.
From the definition of the flux function (Equation~\eqref{eq:Bz}), 
the disk-induced field strength is given by 
\beq
B_{z,d} = \frac{1}{r}\frac{\pd \psi_d}{\pd r}.
\label{eq:Bzd}
\eeq

\subsection{Region I}\label{sec:psid_I}
As we did in Appendix~\ref{sec:BS_append}, we perform the integration in Equation~\eqref{eq:BS_0_I} by expanding the kernel $R(r,r')$ in powers of $r/r'$. 
Applying Equations~\eqref{eq:BS_appendix} and \eqref{eq:azimuth_expand} to region I 
where $r<\rin<r'$, we obtain an equation similar to Equation~\eqref{eq:psid_expand0},
\beq
\psi_{d,\I} = \frac{\pi}{c} \sum_{n=0}^\infty c_n r^{2(n+1)}
\int_\rin^\rout \frac{K_\phi(r')}{r'^{2n+1}}dr'.
\eeq
Below we will assume $\rout \to \infty$ since the contribution 
of the outer boundary to $\psi_{d,\I}$ I is small. 

Inserting the current density given by Equation~\eqref{eq:Kphi_II_sol} 
into the above equation and neglecting the correction factor $\fout$ for the outer boundary,
we have 
\beq
\psi_{d,\I} = \frac{1}{2} B_\infty \rrout \sum_{n=0}^\infty c_n r^{2(n+1)}
\int_\rin^\infty \frac{\fin(r')}{r'^{2n+3}}dr',
\eeq
where the correction factor $\fin$ for the inner boundary is given by Equation~\eqref{eq:fin}.
The integral can be reduced, by introducing a variable $x = (\rin/r)^2$, to
\beq
\int_\rin^\infty \frac{\fin(r')}{r'^{2n+3}}dr'
= \frac{\int_{0}^1 x^n (1-x)^{-1/2}dx}{2r_{\rm in}^{2(n+1)}} 
= \frac{ {B}(n+1,\tfrac{1}{2}) }{2r_{\rm in}^{2(n+1)}},
\eeq
where ${B}(x,y)$ is the Beta function.
It can also be shown with {\it Mathematica} that 
\beq
\sum_{n=0}^\infty c_n {B}(n+1,\tfrac{1}{2}) \pfrac{r}{\rin}^{2(n+1)} 
= 4-4 \sqrt{1- \pfrac{r}{\rin}^2 }.
\eeq
Using these relations, we finally obtain
\beq
\psi_{d,\I} = B_\infty \rrout \left[1- \sqrt{1- \pfrac{r}{\rin}^2 } \right].
\label{eq:psid_I_exact}
\eeq

From Equation~\eqref{eq:Bzd}, we also obtain the strength of the induced vertical field, 
\beq
B_{z,d,\I} = \pfrac{\rout}{\rin}^2 \frac{B_\infty}{\sqrt{1- (r/\rin)^2}}.
\label{eq:Bz_I_exact}
\eeq
Note that $B_{z,\I} \gg B_\infty$ and hence $B_z \approx B_{z,d,\I}$. 
In the limit of $r \ll \rin$, 
Equations~\eqref{eq:psid_I_exact} and \eqref{eq:Bz_I_exact} reduce to 
Equations~\eqref{eq:psid_I_sol} and \eqref{eq:Bz_I_asympt} in the main text, respectively.

\subsection{Region III}\label{sec:psid_III}
Similarly, we perform the integration in Equation~\eqref{eq:BS_0_III} 
by expanding $R(r',r)$ using Equations~\eqref{eq:BS_appendix} and \eqref{eq:azimuth_expand}. Noting that $r'<\rout<r$ for region III, we obtain
\beq
\psi_{d,\I} = \frac{\pi}{c} \sum_{n=0}^\infty c_n r^{-(2n+1)}
\int_\rin^\rout r'^{2(n+1)} K_\phi(r') dr'.
\eeq
Below we will assume $\rin \to 0$ and neglect the contribution 
of the inner boundary to $\psi_{d,\III}$.

Inserting Equation~\eqref{eq:Kphi_II_sol} into the above equation and 
neglecting the inner correction factor $\fin$,
we have 
\beq
\psi_{d,\III} = \frac{1}{2} B_\infty \rrout \sum_{n=0}^\infty c_n r^{-(2n+1)}
\int_0^\rout r'^{2n} \fout(r') dr',
\label{eq:psi_III_expand}
\eeq
where $\fout$ is given by Equation~\eqref{eq:fout}.
By introducing a variable $y = (r/\rout)^2$, 
the integration can be performed as
\beqn
&&\int_0^\rout r'^{2n} \fout(r') dr'
\nonumber \\
&&= \frac{r_{\rm out}^{2n+1}}{2} \int_{0}^1 
\left[  y^{n-\frac{1}{2}} -  y^n + y^n(1-y^n)^\gamma \right] dy 
\nonumber \\
&& = \frac{r_{\rm out}^{2n+1}}{2}\left[
\frac{1}{1+3n+2n^2} + {B}(1+n,1+\gamma)
\right] ,
\label{eq:foutint}
\eeqn
where ${B}(x,y)$ is again  the Beta function.
It can then be shown with {\it Mathematica} that 
\beqn
&&\sum_{n=0}^\infty c_n 
\left[ \frac{1}{1+3n+2n^2} + {B}(1+n,1+\gamma) \right] 
\pfrac{\rout}{r}^{2(n+1)} 
\nonumber \\
&&= \frac{2r}{\rout} \left[ 
{F}\left(-\tfrac{1}{2},-\tfrac{1}{2},1, \tfrac{\rrout}{r^2}\right) -
{F}\left(-\tfrac{1}{2},\tfrac{1}{2},1+\gamma, \tfrac{\rrout}{r^2}\right)
\right],
\eeqn
where ${\rm F}(a,b,c;z)$ is the hypergeometric function. 
Using these relations, $\psi_{d,\III}$ can be written as
\beq
\psi_{d,\III} = \frac{B_\infty r \rout}{2} 
\left[ 
{F}\left(-\tfrac{1}{2},-\tfrac{1}{2},1, \tfrac{\rrout}{r^2}\right) -
{F}\left(-\tfrac{1}{2},\tfrac{1}{2},1+\gamma, \tfrac{\rrout}{r^2}\right)
\right],
\eeq
and from Equation~\eqref{eq:Bzd}, we also obtain the induced vertical field 
\beq
B_{z,d,\III} = \frac{\rout}{r}\left[\frac{2}{\pi}\EllipticE\left(\tfrac{\rrout}{r^2}\right) 
- {F}\left(\tfrac{1}{2},\tfrac{1}{2},1+\gamma, \tfrac{\rrout}{r^2}\right)
\right] .
\eeq

The asymptotic form of  $\psi_{d,\II}$ at $r \gg \rout$ can be easily obtained by dropping the $n \geq 1$ terms in Equation~\eqref{eq:psi_III_expand}, namely,
\beqn
\psi_{d,\III} &\approx& \frac{B_\infty \rrout}{2r} \int_0^\rout \fout(r')dr'
\nonumber \\
&\approx& \frac{B_\infty r_{\rm out}^3}{4r} \left[ 
1+{B}(1,1+\gamma) \right]
\approx 0.425\frac{B_\infty r_{\rm out}^3}{r},
\label{eq:psid_III_exact}
\eeqn
where we have used that Equation~\eqref{eq:foutint}, $c_0 = 1$, 
and ${B}(1,1+\gamma) \approx 0.7$ for the best-fit parameter $\gamma = 0.43$ (see Appendix~\ref{sec:out}). 
The corresponding $B_z$ is 
\beq
B_{z,d,\III} \approx -0.425\pfrac{B_\infty r_{\rm out}}{r}^3.
\label{eq:Bzd_III_exact}
\eeq



\end{document}